\documentclass{aa}
\usepackage{graphics}
\begin {document}
\title{The NASA Astrophysics Data System: Architecture}

\thesaurus{04(04.01.1)}
\author{A. Accomazzi\and G. Eichhorn\and M. J. Kurtz\and C. S. Grant
\and S. S. Murray}
\institute{Harvard-Smithsonian Center for Astrophysics, Cambridge, MA 02138}

\offprints{A. Accomazzi}
\mail{A. Accomazzi}

\date{Received / Accepted}

\titlerunning{}
\authorrunning{A. Accomazzi et al.}

\maketitle

\sloppy

\begin {abstract}
The powerful discovery capabilities available in
the ADS bibliographic services are possible thanks to the design
of a flexible search and retrieval system based on a relational 
database model.
Bibliographic records are stored as a corpus of
structured documents containing fielded data and metadata,
while discipline-specific knowledge is segregated in a set of files 
independent of the bibliographic data itself.
This ancillary information is used by the database management software
to compile field-specific index files used by the ADS search engine 
to resolve user queries into lists of relevant documents.

The creation and management of links to both internal and external 
resources associated with each bibliography in the database
is made possible by representing them as a set of 
document properties and their attributes.  
The resolution of links available from different
locations has been generalized to allow its control through a site- and
user-specific preference database.
To improve global access to the ADS data holdings, a number of 
mirror sites have been created by cloning the database contents
and software on a variety of hardware and software platforms.

The procedures used to create and manage the database and its mirrors 
have been written as a set of scripts that can be 
run in either an interactive or unsupervised fashion.
The modular approach we followed in software development has allowed a high
degree of freedom in prototyping and customization, making our system
rich of features and yet simple enough to be easily modified on a
day-to-day basis.

We conclude discussing the impact that new datasets, 
technologies and collaborations is expected to have on the ADS and
its possible role in an integrated environment of networked
resources in astronomy.

The ADS can be accessed at http://adswww.harvard.edu 

\end{abstract}

\section {\label {intro} Introduction}

The Astrophysics Data System (ADS) Abstract Service was originally
designed as a search and retrieval system offering astronomers and
research librarians sophisticated bibliographic search capabilities. 
Over time, the system has evolved to include full-text scans of 
the scholarly astronomical literature and an ever-increasing 
number of links to resources available from other information
providers, taking full advantage of the
capabilities offered by the emerging technology of the 
World-Wide Web (WWW).

As new data and functionality were incorporated in the ADS,
the design of its system components evolved as well,
driven by the desire to strike a balance between simplicity 
in the operation of the system and richness in its features.
Over time, we favored design approaches promising long-term 
rewards over short-term gains, within the limits allowed
by our resources.
The approach we followed in software development has always been very
pragmatic and data-driven, in the sense that specialized software
components were designed to work efficiently with the existing
datasets, rather than attempting to use general-purpose,
monolithic software packages.

This paper gives an overview of the architecture of
the Astrophysics Data System bibliographic services and discusses
in detail the design of the underlying data structures 
and the implementation of its key software components.
In conjunction with three other ADS papers in this volume,
it is intended to give a complete description of the 
current state and capabilities of the ADS.
An overview of the history and current use of the system is 
given in \cite{OVERVIEW} (OVERVIEW from here on);
details on the datasets in the ADS, their creation and maintenance 
is given in \cite{DATA} (DATA);
a complete description of the ADS search engine and its
user interface is given in \cite{SEARCH} (SEARCH).

Section \ref{creation} discusses the methodological approach
used in the management of bibliographic records,
their representation in the system, and the procedures used
for data exchange with our collaborators.
Section \ref{indexing} describes the structure of
the index files used by the ADS search engine,
the implementation of the procedures that create them,
and the use of discipline-specific knowledge
to improve search results.
Section \ref{properties} details the design and
implementation of general procedures for the creation 
and management of properties associated with bibliographic 
records, and their use in the creation of links to 
internal and external resources.
Section \ref{mirrors} discusses the set of procedures 
used to clone the ADS bibliographic services
to the current mirror sites and the 
level of system independence necessary for their operation.
In section \ref{discussion} we describe how the recent
developments in technology and collaborations among 
astronomical data centers may affect the evolution of
the ADS.

\section {\label {creation} Creation of Bibliographic Records}

The bibliographic records maintained by the ADS project consist of a
corpus of structured documents describing scientific publications.  
Each record is assigned a unique identifier in the system and all
data gathered about the record are stored in a
single text file, named after its identifier.  
The set of all bibliographic records available to the ADS 
is partitioned into four main
data sets: Astronomy, Instrumentation, Physics and Astronomy Preprints
(DATA).  This division of documents into separate groups reflects the
discipline-specific nature of the ADS databases, as discussed in DATA
and section \ref{knowledge}.

Since we receive
bibliographic records from a large number of different sources and in
a variety of formats (DATA), the creation and management of these
records require a system that can parse, identify, and merge
bibliographic data in a reliable way.  In this section we describe the
framework used to implement such a system and some of its
design principles.
Section \ref{methodology} details the methodology behind our
approach.  Section \ref{xml} describes the file format adopted to
represent the bibliographic records.  Section \ref{harvest}
outlines the procedures used to automate data exchange between our
system and our collaborators.
Details about the pragmatic aspects of creating and managing the
bibliographic records are described in DATA.

\subsection {\label{methodology} Methodology}

When the ADS abstract service was first introduced to the astronomical
community (\cite{1993adass...2..132K}), the system was built on
bibliographic data obtained from a single source (the NASA STI
project, also known as RECON) and in a well-defined format
(structured ASCII records).  
The activity of entering these data into the ADS database consisted
simply in parsing the individual records, identifying the different
bibliographic fields in them, and reformatting the contents of these
fields into the ones used in our system.
Bibliographical records were created as text files named after STI's
accession numbers (DATA), which the project used to uniquely identify
records in the system.

As the desire for greater inter-operability with other data services
grew (OVERVIEW), the ADS adopted the
bibliographic code (``bibcode'' from here on) 
as the unique identifier for a bibliographic entry (DATA).
This permitted immediate access to the astronomical databases
maintained by the Strasbourg Data Center (CDS), and allowed
integration of SIMBAD's object name resolution
(\cite{1988alds.proc..323E}) within the ADS abstract service
(OVERVIEW).

As more journal publishers and data centers became providers of
bibliographic data to our project,
a unified approach to the creation of bibliographic records became
necessary.
What makes the management of these records challenging is the fact
that we often receive  data about the same bibliographic entry from
different sources, in some cases with incomplete or conflicting
information (e.g. ordering or truncation of the author list).
Even when the data received is semantically consistent, there may be
differences in the way the information has been represented in the
data file.  For instance, while most journal publishers provide us
with properly encoded entities for accented characters and
mathematical symbols, the legacy data currently found in our databases
and provided to us by some sources only contain plain ASCII
characters.
In other, more subtle and yet significant cases, the slightly
different conventions adopted by different groups in the creation of
bibcodes (DATA) make it necessary to have ``special case'' provisions
in our system that take these differences into account when matching
records generated from these sources.

\begin{figure}
\resizebox{\hsize}{!}{\includegraphics{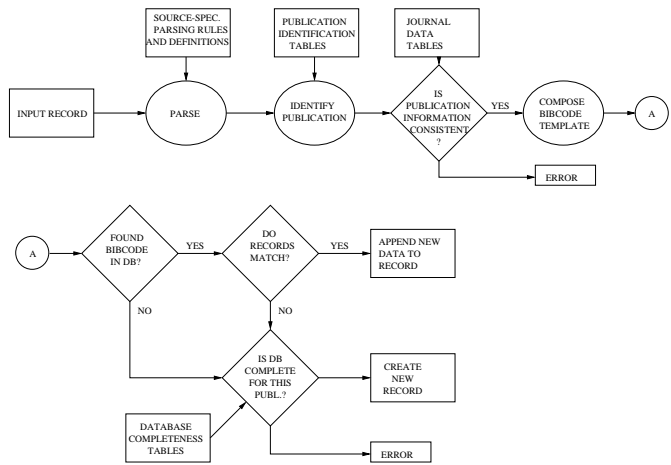}}
\caption[]{Paradigm used for the creation of bibliographic records in the ADS.  }
\label{ADS_architectureF1}
\end{figure}

The paradigm currently followed for the creation of bibliographic
records in our system is illustrated in figure \ref{ADS_architectureF1}.
The different action boxes and tests displayed in the diagram
represent modular procedures, most of which have been implemented as
PERL (\cite{PERL}) software modules.  
More details about each of the
software components can be found in DATA.

As the holdings of the ADS databases have grown over time, additional
metadata about the literature covered in our databases
has been collected and is currently
being used by many of our software modules for a variety of tasks.
Among them it is worth mentioning two activities which are significant
in the context discussed here:

1) Identification of publication sources.  
This is the activity of associating the
name of the publication with the standard abbreviation used to compose
bibliographic codes, and allows us to compute a bibcode for each
record submitted to our system.

2) Data consistency checks.  For all major serials and conference
series in our databases, we maintain tables correlating the volume,
issue, and page ranges with publication dates.  We also have recently
started to maintain ``completeness'' tables describing in analytical
form what range of years or volumes are completely abstracted in our
system for each publication.  This allows us to flag as errors those
records referring to publications for which the ADS has complete coverage,
but which do not match any entry in our system.
The availability of this feature is particularly significant for
reference resolution, as discussed later in this paper.

\subsection {\label{xml} Data Representation}

From the inception of the ADS databases until recently, each
bibliographic record has been represented as a single entity
consisting of a number of different fields (e.g. authors, title,
keywords).  This information was stored in the database as an ASCII
file containing pairs of field names and values.  
While this model has allowed us to keep a structured representation of
each record, over the years its limitations have become apparent.

First of all, the issue of dealing with multiple records referring to
the same bibliographic entry arose.  As previously mentioned, while much of
the information present in these records is the same, certain fields
may only appear in one of them (for example, keywords assigned by the
publisher).  Therefore the capability of managing
bibliographic fields supplied by different sources became desirable,
which could not be easily accomplished with the file format being used.

Secondly, the problem of maintaining ancillary information about a
particular bibliographic entry or even an individual bibliographic
field surfaced.  Information such as the time-stamp
indicating when a bibliographic entry was created or modified, which
data provider submitted it, and what is the identifier assigned to the
record by the publisher can be used to decide how this 
data should be merged into our system or how
hyperlinks to this resource should be created.
Even more importantly, it is often necessary to attach semantic
information to individual records.  For instance, if keywords are
assigned to a particular journal article, it is important to know what
keyword system or thesaurus was used in order to effectively use this
information for document classification and retrieval (\cite{DUBIN99}).

Thirdly, the issue of properly structuring the bibliographic fields had
to be considered.  Some of these fields contain simply plaintext
words, and as such can be easily represented by unformatted
character strings.  Others, however, consist of lists of items
(e.g. keywords or astronomical objects), or may contain structured
information within their contents (e.g. an abstract containing tables
or math formulae).
The simple tagged format we had adopted did not allow
us to easily create hierarchical structures containing 
subfields within a bibliographic field.

Finally, there was the problem of representing relationships among
bibliographical entries (e.g. an erratum referring the original
paper), or among bibliographic fields (e.g. an author corresponding 
to an affiliation). 
While we had been using ASCII identifiers to cross-correlate authors
and affiliations in our records, the adopted scheme was very limited
in its capabilities (e.g. multiple affiliations for an author could
not be expressed using the syntax we implemented).

Given the shortcomings of the bibliographic record representation
detailed above, we recently started reformatted all our
bibliographic records as XML (Extensible Markup Language)
documents.  XML is a markup language which is receiving widespread
endorsement as a standard for data representation and exchange.  
Using this format, a single XML document was created for each
bibliographic entry in our system.  Each
bibliographic field is represented as an XML element,
and may in turn consist of sub-elements (see DATA for an
example of such a file).
Ancillary information about the record is stored as metadata elements
within the document.  Information about an individual field
within the record is stored as attributes of the element representing
it.  Relationships among fields are expressed as links between the
corresponding XML elements.

While it is beyond the scope of this paper to describe the
characteristics that make XML a desirable language for representing
structured documents, we will point out the main reasons why XML was
selected over other formats in our environment.  The reader should
note that most of these remarks not only apply to XML, but also to
its ``parent'' language, SGML (Standard Generalized Markup Language).

XML can be used to represent precise, possibly non-textual information
organized in data structures, and as such can be used as a formal
language for expressing complex data records and their relationships.
In our case, this means that bibliographic fields can be described in
as much detail as necessary.  For instance, the publication
information for a conference proceedings volume can be composed of the
conference title, the conference series name and number, the names of
the editors, the name of the publisher, the place of publication, and the
ISBN number for the printed book.  While all this information has been
stored in the past in a single bibliographic field, the obvious
representation for it is a structured record where items such as
conference title and editors are clearly indentified and tagged.
This allows, among
other things, to properly identify individual bibliographical items
when formatting the record for a particular application (e.g. when
citing a work in an article).

A second important feature which XML offers is the possibility of
representing any amount of ancillary information (the ``metadata'')
along with the actual contents of a
document.  This permits, among other things, to tag bibliographic
records, or even individual fields, with any relevant piece of information.
For instance, an attribute can be assigned 
to the bibliographic field listing a set of keywords
describing what keyword system they belong to.

Other important characteristics of XML are: the adoption of Unicode
(\cite{UNICODE}) for character data representation, allowing uniform
treatment of all international characters and most scientific symbols;
and the support for standard mechanisms for managing complex
relationships among different documents through hyperlinking.

Some of the practical advantages of adopting XML over other SGML
variants simply come from the wide acceptance of the language in the
scientific community as well as in the software industry.
There is currently great interest among the astronomical
data centers in creating interfaces capable of seamlessly exchanging
XML data (\cite{1999AAS...194.8304S,AML}).
It is our hope that as our implementation of an XML-based markup
language for bibliographic data evolves, it can be integrated in the
emerging Astronomical Markup Language (\cite{AML}).
As many of the technologies in the field of document management change
rapidly, it is important for a project of our scope to adopt the ones
which offer the greatest promise of longevity.  In this sense, we feel
that the level of abstraction and dataset independence that XML
imposes on programmers and data specialists is justifies the added
complexity.

\subsection {\label{harvest} Data Harvesting}

Of vital importance to the operation of the ADS is the issue of data
exchange with collaborators, in particular the capability to
efficiently retrieve data produced by publishers and data providers.
The process of collecting and entering new bibliographic records in
our databases has benefitted from three main developments: 
the adoption by all publishers of electronic production systems from
the earliest stages of their publication process; 
the almost exclusive use of SGML and LaTeX as the formats for document
production;
and the pervasive use of the Internet as the medium for data exchange.

An overview of the procedures used to collect bibliographic data in
the daily interactions between ADS staff and data providers is
presented in DATA.
In this section we discuss how the use of automated procedures has
benefitted the activities of data retrieval and entry in the 
operations of the ADS.  
Two approaches are presented: the ``push''
paradigm, in which data is sent from the data provider to the ADS, and
the ``pull'' paradigm, in which data is retrieved from the data
provider.

\subsubsection {\label{push} Data Push}

The ``push'' approach has received much attention since the
introduction of web-based broadcasting technologies in 1997
(\cite{CDF}), to the point that many people consider both push and
web broadcasting to have the same meaning.  Here we refer to the
concept of data ``push'' in its original meaning, i.e. the activity of
electronic data submission to one or more recipients.
The primary means used by ADS users and collaborators to send us
electronic data are: FTP upload, e-mail, and submission through a web
browser (DATA).  
While these three mechanisms are conceptually similar (data is sent
from a user to a computer server using one of several
well-established Internet protocols), the one we have found most
amenable to receiving ``pushed'' data is the e-mail approach. 
This is primarily due to the fact that modern electronic mail 
transport and delivery agents offer many of the features necessary to
implement reliable data delivery, including content encoding, error
handling, data retransmission and acknowledgement.
Additional features such as strong authentication and encryption can
be implemented at a higher level through the use of proper software
agents after data delivery has been completed.
In the rest of the section we describe the implementation of an
email-based data submission service used by the ADS, although the
system operation can be easily adapted to work under other delivery
mechanisms such as FTP or HTTP.

In an attempt to streamline the management of the increasing amount of
bibliographic data sent to us, we have put in place procedures to
automatically filter and process messages sent to an e-mail address
which has been created as a general-purpose submission mechanism.
This activity is implemented by using the procmail filter
package.
Procmail is a very flexible software tool that has been used 
in the past to automatically process
submission of electronic documents by a number of
institutes (\cite{1999adass...8..257B,1996adass...5..451B}).
Our procmail filter has been configured to analyze the input message, verify
its origin, identify which dataset it belongs to, and archive the body
of the message in the proper dataset-specific directory.  Optionally,
the filter can be set up so that one or more procedures are executed
after archival.  
Most of the submissions received this way are simply archived and
later loaded into the databases by the ADS administrators during a
periodic update (DATA).  
Using this paradigm, the email filter allows us to efficiently manage
submissions from different collaborators by enforcing authentication
of the submitter's email address and by properly filing the message
body. 
This procedure is currently used to archive the IAU Circulars and the
Minor Planet Electronic Circulars.

By defining additional actions to be performed after archival of a
submitted e-mail message, automated database updates can be
implemented.  We currently use this procedure to allow automated
submission and updating of our institution's preprint database, which
is currently maintained by the ADS project as a local resource for
scientists working at the Center for Astrophysics.
The person responsible for maintaining the database contents simply
sends a properly formatted email message to the ADS manager account
and an update operation on the database is automatically triggered;
when the updating is completed, the submitter is notified of the
success or failure of the procedure.
We expect to make increasing use of this capability as the electronic
publication time-lines have been steadily decreasing.

\subsubsection {\label{pull} Data Pull}

``Data pull'' is the activity of retrieving data from one or more 
remote network locations.  According to this model, the retrieval
is initiated by the receiving side, which simply downloads the data
from the remote site and stores it in one or more local files.
We have been using this approach for a number of years to retrieve
electronic records made available online by many of our
collaborators.  For instance, the ADS LANL astronomy preprint database
(SEARCH) is updated every night by a procedure that retrieves the
latest submissions of astronomy preprints from the Los Alamos
National Laboratory (LANL) archive, 
creates a properly formatted copy of them in the ADS database, and
then runs an updating procedure that recreates the index files used by
the search engine (section \ref{indexing}).  
This nightly procedure has been running in an unsupervised fashion
since the beginning of 1997.

The pull approach is best used to periodically harvest data that
may have changed.  By using procedures that are capable of
saving and comparing the original timestamps generated by web
servers we can avoid retrieving a network resource unless it has been
updated, making efficient use of the bandwidth and resources
available.   Section \ref{propsoft} discusses the application of these
techniques to the management of distributed bibliographic resources.

\section {\label {indexing} Indexing of Bibliographic Records}

In the classic model of information retrieval (\cite{SALTON83,BC92}), 
the function of a document indexing engine is: 
the extraction of relevant items from the collection of text;
the translation of such items into words belonging to the so-called
Indexing Language (\cite{SALTON83});
and the arrangement of these words into data structures that
support efficient search and retrieval capabilities.
Similarly, the function of a search engine is:
the translation of queries into words from the Indexing Language;
the comparison of such words with the representations of the documents
in the Indexing Language; 
and the evaluation and presentation of the results to the user.

The heterogeneous nature of the bibliographic data entered
into our database (DATA), and the need to effectively deal with the
imprecision in them
led us to design a system where a large set of discipline-specific
interpretations are made. For instance, to cope with the different use of
abstract keywords by the publishers, and to correct possible spelling errors
in the text, sets of words have been grouped together as synonyms
for the purpose of searching the databases. Also, many astronomical
object names cited in the literature are translated in a uniform fashion
when indexing and searching the database to improve recall and
accuracy.

In order to achieve a high level of software portability and database
independence, the decision was made to write general-purpose indexing and
searching engines and incorporate discipline-specific knowledge in a set
of configuration and ancillary files external to the software itself.
For instance, the determination of what parsing algorithm or program
should be used to extract tokens indexed in a particular 
bibliographic field was left as a configurable option to the indexing 
procedure.  
This allowed us, among other things, to reuse the same code for
parsing text both at search and index time, guaranteeing consistency
of results.

The remainder of this section describes the design and implementation
of the document indexing system used by the ADS:
section \ref{overview} provides an overview of indexing procedures; 
section \ref{knowledge} details the organization of the knowledge base used
during indexing;
section \ref{implementation} discusses the implementation of the 
indexing engine.
Details on the search engine and user interface can be found in 
SEARCH.

\subsection {\label {overview} Overview of the Indexing Engine}

The model we followed for providing search capabilities to the ADS
bibliographic databases makes use of data structures commonly referred
to as inverted files or inverted indices (\cite{KNU73,FB92}).  
To allow the implementation of fielded queries, 
an inverted file structure is created for each searchable field,
as described in section
\ref{implementation}.
(In the following we will refer to 
``bibliographic fields'' as the elements composing a
bibliographic record described in the previous section 
--- e.g. authors, affiliations, abstract ---
and ``search fields'' as all the possible searchable entities
implemented in the query interface and described in detail in  
SEARCH
--- e.g. author, exact author, and text).
In general the mapping between search fields and index files is 
one-to-one, while the mapping between inverted files and bibliographic
fields is one-to-many.  
For instance, in our current implementation, the ``author'' index
consists of the tokens extracted from the authors field, while the
``text'' index is created by joining the contents of the following
fields: abstract, title, keywords, comments, and objects.
The complete mapping between bibliographic fields and search fields is
described in section \ref{implementation}.

During the creation of the inverted files, the indexing engine makes
use of several techniques commonly used in Natural Language Processing
(\cite{EF96}) to improve retrieval accuracy and to implement
sophisticated search options.  These transformations provide the
mapping between the input data and the words belonging to the Index
Language. Some of them are described below.

Normalization:  This procedure converts
different morphological variants of a term into a single
format.  The aim of normalization is to reduce redundancy in
the input data and to standardize the format of some particular
expressions.  This step is particularly important when treating data
from heterogeneous sources which may contain textual representations
of mathematical expressions, chemical formulae, astronomical object
names, compound words, etc.  A description of how this is implemented
via morphological translation rules is provided in section \ref{morpho}.

Tokenization:  This procedure 
takes an input character string and returns an array of elements considered
words belonging to the Index Language.  While the tokenization of
well-structured fields such as author or object names is
straightforward, parsing and tokenizing portions of free-text data is
not a trivial matter.  For instance, the decision on how to split into
individual tokens expressions such as ``non-N.A.S.A.'' or 
designations for an astronomical object such as ``PSR 1913+16'' is
often both discipline and context-specific.
To ensure consistency of the search interface and index files, the same
software used to scan text words at search time is used to parse the
bibliographic records at indexing time.  A detailed description of the
text tokenizer is presented in SEARCH.

Case folding: Converting the case of words during indexing is a
standard procedure in the creation of indices and allows the reduction
in size of most index files by removing redundancy in the input data.
For example, converting all text to uppercase both at indexing and
search time allows us to map the strings ``SuperNova,'' ``Supernova,''
and ``supernova'' to the canonical uppercase form ``SUPERNOVA.''
In our implementation the feature of folding case has been set as an
option which can be selected 
on a field-by-field basis, since case is significant in some
rare but important circumstances (e.g. the list of planetary objects).
Details on the treatment of case in fielded queries are discussed
in SEARCH.

Stop words removal: The process of eliminating high-frequency function
words commonly used in the literature also contributes to reduce the
amount of non-discriminating information that is parsed and indexed
(\cite{SALTON83}).
The use of case-sensitive stop words (described in section \ref{stop})
allows us to keep those words in which case alone can discriminate the 
semantics of the expression.

Synonym expansion: By grouping words in synonym classes we can
implement a so-called ``query expansion'' by returning not only the
documents containing one particular search item, but also the ones
containing any of its synonyms.  Using a well-defined set of synonyms
rather than relying on grouping words by stemming algorithms to
perform query expansion provides much greater control in the
implementation of query expansion and can yield a much greater level
of accuracy in the results.  This powerful feature of the ADS indexing
and search engines is described more fully in section \ref{synonyms}.

\subsection {\label {knowledge} Discipline-specific Knowledge Base}

The operation of the indexing engine is driven by a set of ancillary
files representing a knowledge base (\cite{HR83}) which is specific to the
domain of the data being indexed.  This means that in general different
ancillary files are used when indexing data in the different
databases, although in practice much of the metadata used is 
shared among them.

Since the input bibliographies
consist of a collection of fielded entries and each field contains 
terms with distinct and well-defined syntax and semantics, the processing 
applied to each field has to be tailored to its contents.
The following subsections describe the different components of the
knowledge base in use.

\subsubsection {\label{morpho} Morphological Translation Rules}

Morphological translation rules are syntactic operations designed to
convert different representations of the same basic literal expression
into a common format (\cite{SALTON83}).
This is most commonly done with astronomical object names (e.g. 
``M 31'' vs. ``M31''), as well as some composite words (e.g. ``X RAY'', 
``X-RAY'' and ``XRAY'').  The translations are specified as pairs of
antecedent and consequent patterns, and are applied in a
case-insensitive way both at indexing and searching time.  
The antecedent of the translation is usually a POSIX (\cite{POSIX})
regular expression, 
which should be matched against the input data being indexed or
searched.
The consequent is an expression that replaces the
antecedent if a match occurs, and which may contain 
back-references to substrings matched by the antecedent.

The table of translation rules used by the indexing and search
engine uses two sets of replacement expressions for maximum
flexibility in the specification of the translations, one to be used 
during indexing and the other one for searching.  This allows
for instance the contraction of two words into a single expression
while still allowing indexing of the two separate words.  For example,
the expression ``Be stars'' is translated into ``Bestars'' when
searching and ``Bestars stars'' when indexing, so that a search for
``stars'' would still find the record containing this expression.
Note that if we had not used the translation rule described above to
create the compound word ``bestars,'' the
word ``Be'' would have been removed since it is a stop word, and
the search would have just returned all documents containing ``stars.''
The complete list of translation rules currently in use is displayed in
table~\ref{Tmorpho}.

\begin{table*}
\caption[]{Morphological Translation Rules used by the search and 
indexing engines.  The first column contains a sequential rule number.
The second column contains the POSIX regular expression used to 
match input patterns in a case insensitive way;
in this context $\backslash$b represents a word boundary.
The third and fourth columns contain replacement strings used 
when searching and indexing, respectively; most of these 
strings contain backreferences to the patterns matched
by the parenthesized expressions in the antecedent.
The class of expressions matching the different sets of rules can be 
summarized as follows:
1. spectral types of stars;
2. $H_{\alpha}$, $H_{\beta}$, $H_{I}$, $H_{II}$;
3-7. Compound terms;
8. Messier Galaxies;
9. Abell Clusters;
10. NGC Catalog;
11-12. other Catalogs;
13-14. common abbreviations;
15.  supernova 1987A;
16.  english elisions;
17.  french elisions;
18-20. all other elisions;
21-23. chemical symbols and formulae.

}
\label{Tmorpho}
\begin{tabular*}{7.0in}{llll}

\hline
\noalign{\smallskip}

Nr. & Input Pattern & Search replacement & Index replacement
\\
\noalign{\smallskip}
\hline
\noalign{\smallskip}
1.& $\backslash$b(BE$\vert$[OBAFGKMS])(--$\vert$~+)STAR(S?)$\backslash$b	& $\backslash$1STAR$\backslash$3	& $\backslash$1STAR$\backslash$3 STAR$\backslash$3 \\
2.& $\backslash$bH(--$\vert$~+)(ALPHA$\vert$BETA$\vert$I+)$\backslash$b	& H$\backslash$2	& H$\backslash$2 $\backslash$2 \\
3.& $\backslash$bINFRA(--$\vert$~+)RED([A-Z]*)$\backslash$b	& INFRARED$\backslash$2	& INFRARED$\backslash$2 RED$\backslash$2 \\
4.& $\backslash$bRED(--$\vert$~+)SHIFT([A-Z]*)$\backslash$b	& REDSHIFT$\backslash$2	& RED REDSHIFT$\backslash$2 SHIFT$\backslash$2 \\
5.& $\backslash$bT(--$\vert$~+)TAURI$\backslash$b & TTAURI & TTAURI TAURI \\
6.& $\backslash$bX(--$\vert$~+)RAY(S?)$\backslash$b	& XRAY$\backslash$2	& XRAY$\backslash$2 RAY$\backslash$2 \\
7.& $\backslash$bGAMMA(--$\vert$~+)RAY(S?)$\backslash$b	& GAMMARAY$\backslash$2	& GAMMA GAMMARAY$\backslash$2 RAY$\backslash$2 \\
8.& $\backslash$bMESSIER(--$\vert$~+)([0-9])	& M$\backslash$2	& MESSIER $\backslash$2 M$\backslash$2 \\
9.& $\backslash$bABELL(--$\vert$~+)([0-9])	& A$\backslash$2	& ABELL $\backslash$2 A$\backslash$2   \\
10.& $\backslash$bN(--$\vert$~+)([0-9])	& NGC$\backslash$2	& NGC$\backslash$2 \\
11.& $\backslash$b([34]CR?$\vert$ADS$\vert$H[DHR]$\vert$IC$\vert$[MW])(--$\vert$~+)([0-9])& $\backslash$1$\backslash$3	& $\backslash$1$\backslash$3 $\backslash$3 \\
12.& $\backslash$b(MKN$\vert$NGC$\vert$PKS$\vert$PSR[BJ]?$\vert$SAO$\vert$UGC)(--$\vert$~+)([0-9])& $\backslash$1$\backslash$3	& $\backslash$1$\backslash$3 $\backslash$3 \\
13.& $\backslash$bSHOEMAKER(--$\vert$~+)LEVY(--$\vert$~+)([0-9])	& SL$\backslash$3	& SHOEMAKER LEVY $\backslash$3 SL$\backslash$3 \\
14.& $\backslash$bS-Z$\backslash$b	& SUNYAEV-ZELDOVICH		& SUNYAEV-ZELDOVICH	 \\
15.& $\backslash$b1987(--$\vert$~+)A	& 1987A	& 1987A	\\
16.& ([A-Z])'S$\backslash$b	& $\backslash$1	& $\backslash$1	\\
17.& $\backslash$b[DL]'([AEIOUY])	& $\backslash$1	& $\backslash$1	\\
18.& $\backslash$b([A-Z]+[A-Z])'([A-RT-Z])$\backslash$b	& $\backslash$1$\backslash$2	& $\backslash$1 $\backslash$1$\backslash$2 \\
19.& $\backslash$b([A-CE-KM-Z])'([A-Z][A-Z]+)$\backslash$b	& $\backslash$1$\backslash$2	& $\backslash$1$\backslash$2 $\backslash$2 \\
20.& $\backslash$b([A-Z]+[A-Z])'([A-Z][A-Z]+)$\backslash$b	& $\backslash$1$\backslash$2	& $\backslash$1 $\backslash$1$\backslash$2 $\backslash$2 \\
21.& ([A-Z0-9]+)([$\backslash$-$\backslash$+]+)$\backslash$B	& N/A	& $\backslash$1$\backslash$2 $\backslash$1 \\
22.& ([A-Z0-9]*[A-Z])([$\backslash$-$\backslash$+]+)([A-Z0-9]+)	& N/A	& $\backslash$1 $\backslash$1$\backslash$2 $\backslash$1$\backslash$2$\backslash$3 $\backslash$3 \\
23.& ([A-Z0-9]*[0-9])([$\backslash$-$\backslash$+]+)([A-Z][A-Z0-9]*) & N/A & $\backslash$1 $\backslash$1$\backslash$2 $\backslash$1$\backslash$2$\backslash$3 $\backslash$3 \\

\\
\noalign{\smallskip}
\hline
\end{tabular*}
\end{table*}

To avoid the performance penalties associated with matching large
amounts of literal data against the translation rules,
the regular expressions are ``compiled'' into resident RAM
when the ADS services are started,
making the application of regular expressions to the input stream very
efficient (SEARCH).

Despite the extensive use of synonyms in our databases, 
there are cases in which
the words in an input query cannot be found in the field-specific inverted
files.  In order to provide additional search functionality,
two options have been
implemented in the ADS databases, one aimed at improving matching of
English text and a second one aimed at matching of author names.

During the creation of the text and title indices, all words found 
in the database are truncated to their stem according to the 
Porter stemmer algorithm (\cite{HAR91}).
Those stems that do not already appear in the text
and title index are added to the index files and point to the 
list of terms that generated the stem.  Upon searching the database
and not finding a match, the search engine proceeds to apply the same
stemming rules to the input term(s) and then repeat the search.
Thus word stemming is used as a ``last-resort'' measure in an attempt
to match the input query to a group of words that may be related to it.
For searches that require an exact match, no stemming of the input 
query takes place.  The limited use of stemming techniques during indexing
and searching text in the ADS system derives from the observation
that these algorithms only allow minor improvements in the selection
and ranking of search results (\cite{HAR91,XC98}).

To aid in searches on author names, the option to match 
words which are phonetically similar was added in 1996 and is
currently available through one of the ADS user interfaces.
In this case, a secondary inverted file consisting of the different 
phonetic representations of author last names allows a user to generate
lists of last names that can be used to query the database.
Two phonetic retrieval algorithms have been implemented, based on the
``soundex'' (\cite{GADD88}) and ``phonix'' (\cite{GADD90})
algorithms.

\subsubsection {\label {synonyms} Synonym Expansion}

A variety of techniques have been used in information retrieval
to increase recall by retrieving documents containing not only
the words specified in the query but also their synonyms (\cite{EF96}).
By grouping individual words appearing in a bibliographic database
into sets of synonyms, it becomes possible to use this information
either at indexing or searching time to perform a so-called 
``synonym expansion.''

Typically, this procedure has been used as an alternative to text
stemming techniques to automatically search for
different forms of a word (singular vs. plural, name vs. adjective, 
differences in spelling and typographical errors).  
However, since the specification of the synonyms is database- and 
field-specific, our paradigm has allowed us to easily extend the use
of synonyms to other search fields such as authors and planetary
objects (SEARCH). 
Additionally, during the creation of the text synonym groups 
we were able to incorporate discipline-specific knowledge
which would otherwise be missed.  In this sense, the use of synonym
expansion in ADS adds a layer of semantic information that can be 
used to improve search results.
For instance, the following list of
words are listed as being synonyms within the ADS:

\begin{verbatim}

circumquasar		
miniquasar
nonquasar
protoquasars
qso
qsos
qsr
qsrs
qsrss
qss
quarsars
quasar
quasare
quasaren
quasargalaxie
quasargalaxien
quasarhaufung
quasarlike
quasarpaar
quasars
quasers
quasistellar

\end{verbatim}

During indexing and searching, by default  
any words which are part of the same synonym group
are considered to be ``equivalent'' for the purpose of finding
matching documents.  Therefore a title search for ``quasar'' will
also return papers which contained the word ``quasistellar'' in their
title.  Of course, our user interface allows the user to disable 
synonym expansion on a field-by-field as well as on a word-by-word basis.

It is the extensive work that has gone into compiling such a list
that makes searches in the ADS so powerful.  To give
an idea of the magnitude of the task, it should suffice to say that
currently the synonyms database consists of
over 55,000 words grouped into 9,266 sets.  Over the years, 
the clustering of terms in synonym groups has incorporated data from
different sources, including the Multi-Lingual supplement to the 
Astronomy Thesaurus (\cite{1995VA.....39Q.272S}).

Despite the fact that the implementation of query expansion
through the use of synonyms illustrated above 
has shown to be an
effective tool in searching and ranking of results, we are currently
in the process of reviewing the contents and format of the synonym
database to improve its functionality. 
First of all, as we have added more and
more bibliographic references from historical and foreign sources, the
amount of non-English words in our database has been slowly but
steadily increasing.  As a result, we intend to merge the proper
foreign language words with each group of English synonyms in a
systematic fashion (\cite{OARD97,GR98}).

Secondly, we intend to review and correct the current foreign words in
our synonym classes to include, where appropriate, their proper
representation according to the Unicode standard (\cite{UNICODE}), 
which provides
the foundation for internationalization and localization of textual
data.  By identifying entries in our synonym file that were created
by transliterating words that require an expanded character
set into ASCII, 
we can simply add the Unicode representation of the word to the
synonym group, therefore ensuring that both forms will be properly
indexed and found when either form is used in a search.

Finally, we are implementing a more flexible group structure for the
synonyms which allows us to specify hierarchical groupings and
relationship among groups rather than simple equivalence among
words.  This last feature allows us to effectively implement the
use of a limited thesaurus for search purposes (\cite{MILL97}).
Instead of simply grouping words together in a flat structure as
detailed above, we first create separate groups of words, each 
representing a distinct and well-defined concept.  
Words representing 
the concept are then assigned to one such groups and are considered
``equivalent'' instantiations of the concept.  A word can only belong 
to one group but groups can contain subgroups,
representing instances of ``sub-concepts.''
The following XML fragment shows how grouping of synonyms is
being implemented under this new paradigm:

\begin{verbatim}

<syngroup id="00751">
<subgroup rel="instanceof">00752</subgroup>
<subgroup rel="instanceof">00753</subgroup>
<subgroup rel="instanceof">00754</subgroup>
<subgroup rel="instanceof">00755</subgroup>
<subgroup rel="oppositeof">00756</subgroup>
<syn>qso</syn>
<syn>qsos</syn>
...
<syn>quasistellar</syn>
<syn lang="de">quasare</syn>
<syn lang="de">quasaren</syn>
<syn lang="de">quasargalaxie</syn>
<syn lang="de">quasargalaxien</syn>
</syngroup>

<syngroup id="00752">
<syn>circumquasar</syn>
<syn>circumquasars</syn>
</syngroup>

<syngroup id="00753">
<syn>miniquasar</syn>
<syn>miniquasars</syn>
<syn>microquasar</syn>
<syn>microquasars</syn>
</syngroup>

<syngroup id="00754">
<syn>protoquasar</syn>
<syn>protoquasars</syn>
</syngroup>

<syngroup id="00755">
<syn>quasar cluster</syn>
<syn>quasar clusters</syn>
<syn lang="de">quasarh&auml;ufung</syn>
<syn lang="de">quasarh&auml;ufungen</syn>
</syngroup>

<syngroup id="00756">
<syn>nonquasar</syn>
<syn>nonquasars</syn>
</syngroup>

<syngroup id="01033">
...
<subgroup rel="instanceof">00755</subgroup>
...
<syn>cluster</syn>
<syn lang="de">h&auml;ufung</syn>
...
</syngroup>

\end{verbatim}

The new approach allows a much more sophisticated implementation of 
query expansion through the use of synonyms.   Some of its
advantages are:

1) Hierarchical subgrouping of synonyms: every group may contain
   one or more subgroups representing ``sub-concepts'' related
   to the group in question.
   Currently the two relations we make use of are the ones representing
   instantiation and opposition.
   This capability allow us to break down a particular concept at 
   any level of detail, grouping synonyms at each level and then
   ``including'' subgroups as appropriate.

2) Multiple group membership: each subgroup may be an instance
   of one or more synonym groups.  For instance, the synonyms
   ``quasarh\"aufung'' and ``quasar cluster'' are in a subgroup that
   belongs to both the ``qso'' and the ``cluster'' groups.

3) Use of multi-word sequences in synonym groups: in certain cases,
   individual words referring to a concept correspond to a 
   sequence of several words in other languages or context.
   Allowing declarations of multi-word synonyms enables us to correctly
   identify most terms.

4) Multilingual grouping: words belonging to a language other than
   English are tagged with the standard international identifier for that
   language.  This permits us to use the synonyms in a context 
   sensitive way, so that if the same word were to exist in two
   languages with different meanings, the proper synonym group
   would be used when reading documents in each language.

The synonym database described above is used at indexing time to
create common lists of document identifiers for words belonging to the
same synonym group or any of its subgroups.  
The effect of this procedure is that when use of synonyms is enabled,
searches specifying a word that belongs to a synonym group
will result in the list of records containing that word as
well as any other word in the synonym group or its subgroups.
In the example given above, a search for ``qso'' would have
listed all documents containing ``qso,'' its other synonyms,
as well as subgroup members such as ``miniquasar'' and
``protoquasar.''  On the other hand, a search for ``miniquasar''
would have only returned the list of documents containing 
either ``miniquasar'' or ``microquasar,'' narrowing significantly
the search results.

\subsubsection {\label {stop} Stop Words}

A number of words considered ``irrelevant'' with respect to the searches
of the particular field and database at hand are ignored during indexing
and searching.  These words (commonly referred to as ``stop words'') 
consist primarily of terms used in the English language with great frequency, 
as well as adverbs, prepositions and any other words not carrying a 
significant meaning when used in the context under consideration
(\cite{SALTON83}).
Such words are removed both at indexing and searching time,
decreasing the number of irrelevant searches and disregarding search 
terms that would not yield significant results.

The use of both case-sensitive
and case-insensitive stop words during indexing allows us to single out 
those instances of terms that may have different meanings depending on 
their case.  For instance, the words ``he'' and ``He'' usually represent 
different concepts in the scientific literature (the second one being the 
symbol for the element Helium).  
By selectively eliminating all instances of ``he,'' when indexing the
bibliographies, we stand a good chance that the remaining instances of 
the word refer to the element Helium.

The effort currently underway to create a structured synonym database
will be used to group and maintain the list of stop words in use.
By simply clustering stop words in synonym groups and properly tagging
the group as containing stop words, we can use the same software
that is currently being developed to create and maintain the list of
synonyms in our database.  An example of the resulting records is 
shown below:

\begin{verbatim}

<syngroup id="00037" type="stop">
<!-- he is used in case-sensitive way to avoid 
 removing "He" (element helium) from index -->
<syn case="mixed">he</syn> 
<syn>she</syn>
<syn lang="de">er</syn>
<syn lang="de">sie</syn>
<syn lang="fr">il</syn>
<syn lang="fr">elle</syn>
<syn lang="es">&eacute;l</syn>
<!-- as above, but without proper accenting -->
<syn lang="es">el</syn>
<syn lang="es">ella</syn>
<syn lang="it">lui</syn>
<syn lang="it">lei</syn>
</syngroup>

\end{verbatim}

This paradigm allows us to treat stop words as a special case of
synonyms (which are identified by the indexing and search engines
as being of type ``stop'').

\subsection {\label {implementation} The Indexing Engine}

General-purpose indexing engines and relational
databases were used as part of the abstract service in its first
implementation
(\cite{1993adass...2..132K}), but they were eventually dropped
in favor of a custom system as the desire for better performance
and additional features grew with time (\cite{1995adass...4...36A}),
as is often necessary in the creation of discipline-specific
information retrieval systems (\cite{VR79}).
The approach used to implement the data indexing portion of the
database can be considered ``data-driven'' in the sense that
parsing, matching and processing of input text data is controlled
by a single configuration file (described below)
and by the discipline-specific files described in section 
\ref{knowledge}.

The inverted files used by the search engine are the products of a
pipeline of data processing steps that has evolved with time.
To allow maximum flexibility in the definition 
of the different processing steps,
we have found it useful to break down the indexing
procedure into a sequence of smaller and simpler tasks that
are general enough to be used for the creation of all the files
required by the search engine.
A key design element which has helped generalize the indexing process is
the use of a configuration file which describes all the 
field-specific processing necessary to create the index files.
The configuration file currently in use is displayed in table 
\ref{Tindexconfig}.
For each search field listed in the table, 
an inverted file structure is created by the indexing engine.

\begin{table*}
\caption[]{Configurable parameters used by the indexing engine.
The first column lists the searchable fields to be indexed.
The second column lists the XML elements whose contents are used to
create each field's occurrence table; 
note that since the author search field is derived
from the exact author index, no elements are listed for it.
The third column shows whether the contents of the field are
modified through the use of morphological translation rules.
The fourth column lists the name of the procedure used to parse the
field contents into individual tokens to be indexed. 
These procedures currently are: get\_children, which tokenizes an XML
fragment by extracting its subelements from it; text\_parser, which
tokenizes input text as described in the SEARCH paper;
and parse\_authors, which reduces all author names to the canonical
forms used by ADS.
The last three columns show whether stop words, case folding, and
synonym grouping is used during indexing of each search field.
Note that because of the lack of a common set of keywords used
throughout the database, keyword searches are currently disabled in
our standard query interface.
}
\label{Tindexconfig}
\begin{tabular*}{7.0in}{lp{2.in}lllll}

\hline
\noalign{\smallskip}

Search Field & Bibliographic Fields &
Translation & Tokenizer & Stop words &
Case folding & Synonyms

\\
\noalign{\smallskip}
\hline
\noalign{\smallskip}
exact author & $<$AUTHORS$>$	& 		no	& 	get\_children	& no	& 	yes	& 	no \\
title	& 	$<$TITLE$>$ 	&		yes	& 	text\_parser	& yes	& 	yes	& 	yes \\
text	& $<$TITLE$>$, $<$ABSTRACT$>$, $<$KEYWORDS$>$, $<$OBJECTS$>$, $<$COMMENTS$>$ & yes  & text\_parser	& yes	& 	yes	& 	yes \\
keyword 	& $<$KEYWORDS$>$		& no 	& 	get\_children	& no		& yes		& no  \\
object 	& $<$OBJECTS$>$		& 	yes	& 	get\_children	& no		& no		& yes  \\

author	& (exact author occurrence table) &	no	& 	parse\_authors	& no		& yes		& yes \\
\\
\noalign{\smallskip}
\hline
\end{tabular*}
\end{table*}

The first step performed by the indexing software is the creation of a
list containing the document identifiers to be indexed.  This usually
consists of the entire set of documents included in a particular
database but may be specified as a subset of it if necessary (for
instance when creating an update to the index, see section \ref{update}).
The list of document identifiers is then given as input to an ``indexer''
program, which proceeds to create, for each search field, an inverted file
containing the tokens extracted from the input documents
and the document identifiers (bibcodes) where such words occur.
(In the following discussion we will refer to the tokens extracted by
the indexer simply 
as ``words,'' although they may not be actual words in the common
sense of the term.  For instance, during the creation of the author
index, the ``words'' being indexed are author names.)
After all the inverted files have been created, each one of them is
processed by a second procedure which generates two separate files used by
the search engine: an ``index'' file, containing the list of
words along with pointers to a list of document identifiers, and a
``list'' file, containing compact representations of the lists of
document identifiers corresponding to each word.

The following subsections describe the procedures used during the
different indexing steps: 
section \ref{invfiles} details the creation of the inverted files; 
section \ref{indexfiles} describes the creation of the index and list files; 
section \ref{update} describes the procedures used to update the index and
list files;
section \ref{indexsummary} discusses some of the advantages and
shortfalls of the implemented indexing scheme.

\subsubsection {\label {invfiles} Creation of Inverted Files}

An inverted file (\cite{VR79,FB92}) is a table consisting of
two columns: the first column contains the instances of words
belonging to the indexing language, and the 
second column contains the list of document identifiers in which those
words were found.  
The transformation of a document into its indexing language is
performed in the following steps:

1)   parsing of the document contents and extraction of all the
	bibliographic elements needed for the creation of one
	or more search fields;
2)   joining of bibliographic elements that should be indexed together
	to produce a list of strings;
3)   application of translation rules (if any) to the list of strings;
4)   itemization of the list of strings into an array of words to be
	indexed;
5)   removal of stop words from the list of words to be indexed
	(either case sensitively or insensitively);
6)   folding of case for each of the words (if requested);
7)   creation or addition of an entry for each word in a hash table
	correlating the word indexed with the document 
	identifiers where it appears.

The indexer keeps a separate inverted file for each set of indexing
fields to be created (see table \ref{Tindexconfig}, column 1).
Each inverted file is simply implemented as a sorted
ASCII table, with tab separated columns.
Given the current size of our databases, the creation of
these tables takes place incrementally.
A pre-set number of documents is
read and processed by the indexer, 
an occurrence hash table for these documents is
computed in memory, and an ASCII dump of the hash is then written
to disk file as a set of keys (the words being indexed) 
followed by a list of document identifiers containing such words.
The global inverted file is then created by simply joining the
partial inverted files using a variation of the standard UNIX join command.

Once the occurrence tables for the primary search fields listed in
table \ref{Tindexconfig} have been created,
a set of derived fields are computed if necessary.  
Currently this step is used to
create the ``authors'' occurrence table from the ``exact authors'' one
by parsing and formatting entries in it so that all names are reduced
to the forms ``Lastname, F'' (where ``F'' stands for the first name
initial) and ``Lastname.''  This allows efficient searching for the
standard author citation format.

\subsubsection {\label{indexfiles} Creation of the Index and List Files}

After all the primary and derived inverted files have been
generated, a separate program is used to produce for each table
two separate files which are used by the search engine:
an inverted index file (here simply called ``index'' file) 
and a document list file (``list'' file, see \cite{SALTON89}).
The index file is an ASCII table which contains 
the complete list of words appearing in the inverted file and 
two sets of numerical values associated with it, the first set
is used for exact word searches, the second one for synonym
searches.  The list file is a binary file containing blocks of
document identifiers in which a particular word was found.
Each set of numerical values specified in the index file consists of:
the relative ``weight'' of the word (or group of synonyms) 
in the database, as defined below;
the length of the group of document identifiers in the list file,
in bytes;
the position of the group of document identifiers in the list file,
defined as the byte offset from the beginning of the list file.

The value chosen to express the weight $W(w)$ of a word
$w$ is a variation of the inverse document frequency
(\cite{SALTON88}): 

	$$W(w) = K \times \log_{10} {N / df(w)}$$

where $K$ is a constant, $N$ is the total number of documents in the
database, and $df(w)$ is the document frequency of the word $w$,
i.e. the number of documents in which the word appears (\cite{SALTON88}).
The choice of a suitable value for the constant $K$ (currently set to
$K = 10^4$) allows the indexing and search engine  to perform most of
the operations in integer arithmetic.
To avoid performing slow log computations during the creation of
the index files, the function that maps $df(w)$ to $W(w)$ 
is cached in an associative array so that when repeated 
integer values of $df(w)$ are encountered, the pre-computed values are
used.

\begin{figure}
\resizebox{\hsize}{!}{\includegraphics{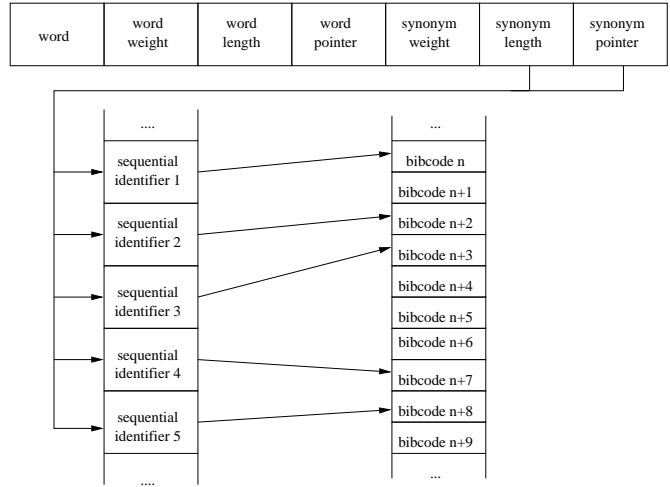}}
\caption[]{ Implementation of the inverted file structure makes use of
multiple lookups for efficiency.  This diagram describes the sequence
of steps performed by the search engine to resolve a fielded query
with synonym expansion enabled. First the query word is found in the
index file which contains a sorted list of strings, and the address
and length of  sequential identifiers corresponding to the word are
read. Then the block of sequential identifiers are read from the list
file and are individually resolved into their corresponding
bibcodes. }
\label{ADS_architectureF2}
\end{figure}

The document identifiers which are stored in the list files are 32-bit
integers (from here on called sequential identifiers) 
corresponding to line numbers in the list of bibliographic
codes which have been indexed.  The search engine resolves all queries
on index files by performing binary searches on the words appearing in
the index file, then reading the corresponding list of sequential
identifiers in the list files, combining results, and
finally resolving the sequential identifiers in bibcodes 
(see figure \ref{ADS_architectureF2}).

The procedure for the creation of the index and list files
reads the inverted file associated with each search field and performs
the following steps:

1) read all entries from the list of document identifiers
  (bibcode list) and create a hash table associating each bibcode
  with its corresponding sequential identifier;

2) if synonym grouping is to be used for this field, read the
  synonym file for this field and create a hash table associating each
  entry in the synonym group with the word with the highest frequency in
  the group;

3) for each word in the inverted file translate the list of
  bibcodes associated with it into the corresponding 
  list of integer line numbers, and mark word as being processed;

4) if word belongs to a group of synonyms, sequentially find and
  process all other words in the same group, marking them as processed,
  then iteratively process all words in any of the subgroups until 
  nesting of subgroups is exhausted; if no synonyms are in use, the
  same procedure is used with the provision that the group of synonyms
  is considered to be composed only of the word itself;

5) join, sort and unique the lists of sequential identifiers 
  for all the words in the current group of synonyms;

6) write to the list file the sorted list of sequential
  identifiers for each word
  in the group of synonyms, followed by the
  cumulative list of sequential identifiers for the entire group of
  synonyms;

7) for each word in the group of synonyms, write to the index file an
  entry containing the word itself and the two sets of numerical
  values (weight, length, and offset) for exact word and synonym searches.

\begin{figure}
\resizebox{\hsize}{!}{\includegraphics{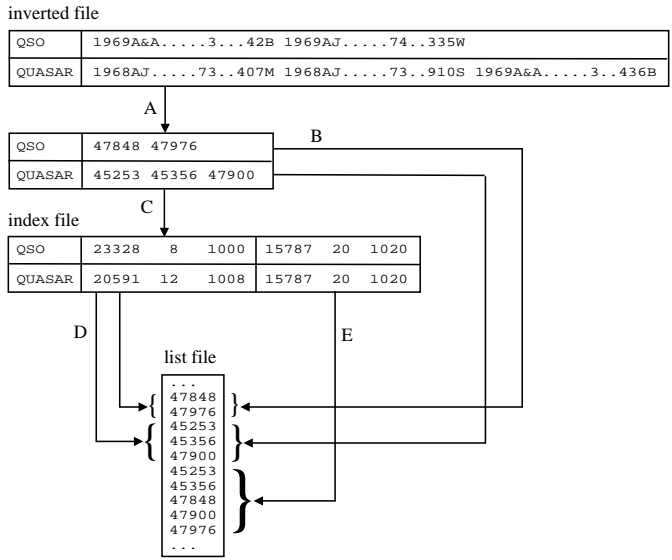}}
\caption[]{ Creation of the text index and list file from the inverted
file.  First the bibliographic codes are translated into sequential
identifiers (A).  Then the list file is created by concatenating
``blocks'' of sequential identifiers for each word  and each group of
synonyms in the inverted file (B), and the index file is created by
storing the list of words, weights, and pointers to these blocks of
sequential identifiers (C). To retrieve the list of documents
containing a word or any of its synonyms, the search engine searches
the index file and then reads the block of identifiers for either
simple word searches (D) or synonym searches (E). }

\label{ADS_architectureF3}
\end{figure}

Figure \ref{ADS_architectureF3} illustrates the creation of entries for two words in the
``text'' index and list file from the text inverted file.

\subsubsection {\label {update} Index Updates}

The separation of the indexing activity into two separate parts offers
different options when it comes to updating an index.  New documents
which are added to the database can be processed by the indexer and
merged into the inverted file quickly, and a new set of index
and list files can then be generated from it.  
Similarly, since the synonym grouping is performed after the creation
of the inverted files, a change in the synonym database can be
propagated to the files used by the search engine by recreating the 
index and list files, avoiding a complete re-indexing of the database.

Despite the steps that have been taken in optimizing the code used
in the creation of the index and list files from the occurrence 
tables, this procedure still takes close to two hours to complete
when run on the complete set of bibliographies in the astronomy database
using the hardware and software at our disposal.
In order to allow rapid and incremental updating of the index and list
files, a separate scheme has been devised requiring only in-place 
modification of these files rather than their complete re-computation.

During a so-called ``quick update'' of an operational set of index
files used by the search engine, a new indexing procedure is run on
the documents that have been added to the database since the last full
indexing has taken place.  The indexing
procedure produces new sets of incremental index and list files 
as described above,
with the obvious difference that these files only contain words that 
appear in the new bibliographic records added to the database.
A separate procedure is then used to merge the new set of index and list files
into the global index and list files used by the operational search 
engine, making the new records immediately available to the user.  
The procedure is implemented in the following steps (see figure
\ref{ADS_architectureF4}):

\begin{figure}
\resizebox{\hsize}{!}{\includegraphics{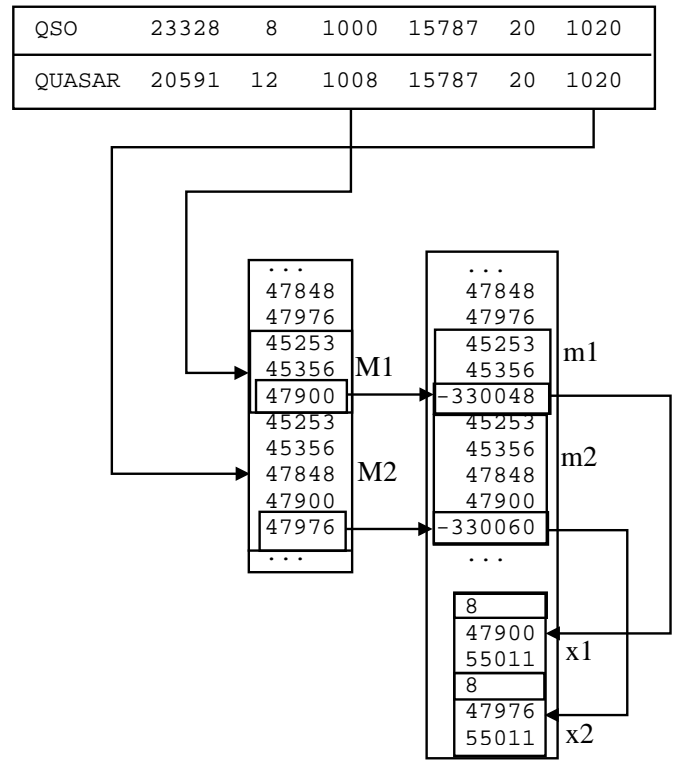}}
\caption[]{ Modification of a list file by a ``quick update:'' the
main blocks corresponding to a word present in the incremental index
(M1 and M2) are modified by the insertion of a pointer at their end
and by extension blocks at the end of the index (x1 and x2). }

\label{ADS_architectureF4}
\end{figure}

1) Compute new sequential identifiers for the list of bibcodes in the
   incremental index by adding to each of them the number of entries
   in the operational bibcode list.  This guarantees that the mapping
   between bibcodes and sequential identifiers is still unique after
   the new bibcodes have been merged into the operational index.

2) Append the list of sequential identifiers found in the incremental
   list file to the operational list file.  In the case of identifiers
   corresponding to a new entry in the index file, their block of 
   values is simply appended to the end of the operational list file.
   In the case of identifiers corresponding to an entry already
   present in the operational index file, the original list of 
   identifiers (``main block'') needs to be merged with
   the new list of identifiers.
   In order to avoid clobbering
   existing data in the operational list file, the list of identifiers
   from the incremental index is appended to the end of the global
   list file, creating an extension of the main blocks of identifiers
   that we call an ``extension block.''
   To accomplish the linking between main and extension blocks, 
   the last sequential identifier in a main block is overwritten
   with a negative value representing the corresponding extension
   block's offset from the  beginning of the list file (except the 
   change in sign).
   An extension block contains as the first integer value the size of
   the extension block in bytes, followed by the last identifier read
   from the main block in the list file, followed by the sequential
   identifiers from the incremental index (see fig. \ref{ADS_architectureF4}).
   When the search engine finds a negative
   number as the last document identifier value, it will seek to the 
   specified offset, read a single integer entry corresponding to the
   number of bytes composing the extension block, and then proceed to
   read the specified number of identifiers.
   Note that because of the way extension blocks are created, the list
   of sequential identifiers created by concatenating the entries in
   the extension block to the entries in the main block is always sorted.

3) For each entry in each incremental index file, determine if a
   corresponding entry exists in the operational index file.
   If an entry is found, no modification
   of the index file is necessary, otherwise the index file is
   updated by inserting the entry in it.  The values of the weights
   and offsets are corrected by taking into account the total number
   of documents in the operational index and the size of the list file.

\subsubsection {\label{indexsummary} Remarks on the Adopted Indexing Scheme}

One of the advantages of using separated index and list files is that
the size of the files that are accessed most frequently by the
search engine (the list of bibcodes and the index files) is kept small so
that their contents can be loaded in random access memory
and searched efficiently (SEARCH).  
For instance, the size of the text index file for the astronomy
database is approximately 16 MB, and once the numerical entries are
converted into binary representation when loaded in memory by the search 
engine, the actual amount of memory used is less than 10 MB.

The use of integer sequential identifiers in the list files
allows more compact storage of the document identifiers as well as 
implementation of fast algorithms for merging search results (since
all the operations are executed in 32-bit integer arithmetic rather
than having to operate on 19-character strings).
For instance, recent indexing of the ADS astronomy database
produces text inverted files which have sizes approaching 
500 MB, while the size of the text list file is about 140 MB.

The choice of a word weight which is a function of only the 
document frequency allows us to store word weights as part of the
index files.  It has been shown that a better measure for the 
relevance of a document with respect to a query word is obtained
by taking into account both the document frequency $df$ and the
term frequency $tf$, defined as the frequency of the word in each
document in which it appears (\cite{SALTON88}), normalized
to the total number of words in the document.  The reasoning
behind this is that a word occurring with high relative frequency
in a document and not as frequently in the rest of the database is
a good discriminant element for that document.
Although we had originally envisioned incorporating document-specific
weights in the list files to take into account the relative term
frequency of each word, we found that little improvement was 
gained in document ranking.
This is probably due to the fact that the collection of documents
in our databases is rather homogeneous as far as document length
and characteristics are concerned.  Eventually the choice was
made to adopt the simpler weighting scheme described above.

The procedures used to create the inverted files can scale well
with the size of the database since the global inverted file is
always created by joining together partial inverted files.
This allows us to limit the number of hash entries used by the indexer
program during the computation of the inverted files. 
According to Heap's law (\cite{HEA78}), and as verified experimentally 
in our databases, a body of $n$ words typically generates a vocabulary of 
size $V = K n^{\beta}$ where $K$ is a constant and $\beta \approx 0.4 -
0.6$ for English text (\cite{NAV98}).
Since the size of the vocabulary $V$ corresponds to the number of 
entries in a global hash table used by the indexing software, we see
that an ever-increasing amount of hardware resources would be 
necessary to hold the vocabulary in memory; our choice of a partial
indexing scheme avoids this problem.
Furthermore, the incremental indexing model is quite suitable to 
being used in a distributed computing environment where different 
processors can be used in parallel to generate the partial inverted
files, as has been recently shown by \cite{KITA97}.

The procedures used to create the list and index files make use of
memory sparingly, so that processing of entries from the occurrence
tables is essentially sequential. 
The only exception to this is the handling of groups of synonyms.
In that case, the data structures used
to maintain the entries for the words in the current synonym group are
kept in memory while the cumulative list of sequential identifiers for
the entire group is built.  The memory is released as soon as
the entries for the current synonym group are written to the list and
index files.

\section {\label{properties} Management of Bibliographic Properties}

By combining bibliographic data and metadata available from several sources
in a single database and by maintaining a list of what properties and
resources are available for each bibliography, the ADS system allows
users to formulate complex queries such as: ``show me all the papers
that cite any paper ever written about the object M87 and the subject
`globular clusters' and which are available online as full-text
documents.''  This query is possible thanks to the
collection and fusion of data from several sources:

1) The astronomical object databases, which maintain a collection of
object names and bibliographies in which they appear.  This search
is performed through a peer-to-peer network connection with the 
SIMBAD (\cite{1988alds.proc..323E})
and NED (\cite{1988alds.proc..335H}) 
database servers, as described in OVERVIEW and SEARCH.  
This first step allows us to find the set of bibliographies on M87.

2) The ADS abstract service indices, which allow a search of all
astronomical papers containing the words ``globular cluster'' or their
synonyms.  This part of the search is performed by the ADS 
search engine and makes use of the local files generated by 
indexing the bibliographic databases as described in section
\ref{indexing}.
This step allows us to discard any bibliographic
entry which does not contain the words ``globular cluster'' in its
text index.

3) The list of citations in the ADS databases, which maintain updated 
lists of astronomical papers and any paper referenced in them.
This allows us to look up the list of papers that have cited 
the selected bibliographic entries, and then proceed to join the
results.

4) The list of papers available electronically from either the 
astronomical journal publishers or the ADS article service,
both of which provide access to full-text articles online.

The query given above illustrates how knowing whether a particular
bibliographic entry possesses a particular property (e.g. whether it
has been cited) and what values may be associated with that property
(e.g. the list of citing papers) can be used as a method for selection
and ranking of query results.  Additionally, the availability
of remote resources for a particular bibliographic entry can be
described as being one of its properties, which in turns allows 
an additional filtering of the result lists.

As new data regarding a bibliographic entry become available, its
record is updated in the ADS database by merging the new information
with the existing entry and possibly by updating its relevance within
the database and its relation with respect to other internal and
external resources.  For instance, when a new paper is published which
references an existing bibliography, the record for the latter paper
needs to be updated by establishing a link between the two
papers; at the same time, the ``citation relevance measure'' for the
paper, computed as the number of times the paper was cited in the
literature, also needs to be updated.

The procedures used in the creation and management of bibliographic 
properties (simply called ``properties'' from here on) in the ADS 
databases are a result of the need for managing resources 
related to bibliographies which may or may not be available locally.
The main characteristics of the property sets as defined in our system
can be summarized in the following list:

1) Some properties simply denote the fact that an entry belongs to
a certain dataset (e.g. whether a paper is refereed or not), others
may have values associated with them (e.g. ``is available online
electronically'' will have as its value the URL of the full-text
paper).  In general, the knowledge of whether an entry in the database
has a certain property allows the search engine to select it for
further consideration when executing a database query, while the
value(s) assumed by this property do not need to be taken into account
until later.

2) The lists of bibliographic identifiers and their properties may be
defined as being either ``static'' or ``dynamic.''  Static properties are
those that once defined do not change in time (e.g. whether a paper is
refereed), while dynamic properties may change their value with
time (e.g. the list of citations for a paper).

3) Some properties may depend on each other (e.g. references and
citations), hence the creation and updating order for these properties
is significant.

Currently the ADS has defined a set of 21 different properties which
are applicable to its bibliographies.  Some of them are listed in
table \ref{Tcodes}.

\begin{table*}
\caption[]{Examples of bibliographic properties defined in the ADS
and their possible values.
}
\label{Tcodes}
\begin{tabular*}{7.0in}{lp{0.6\linewidth}p{0.25\linewidth}}

\hline
\noalign{\smallskip}

Name &
Explanation &
Value(s)

\\
\noalign{\smallskip}
\hline
\noalign{\smallskip}
associated & 	one or more associated bibliographic records exist for
	this entry
	(e.g. erratum or papers published as part of a series) &
	bibcodes of papers associated with bibliographic entry \\
citation &	bibliographic entry has been cited by one or more
	papers in the ADS &
	bibcodes of papers citing bibliographic entry \\
data	&	bibliographic entry has electronic data tables
	published with it &
	URLs of data tables \\
electronic &	a full-text electronic article exist for this
	bibliographic entry & 
	URL of electronic journal article \\
ocr	&	abstract of bibliographic entry was generated by
	Optical Character Recognition programs &
	N/A \\
refereed &	bibliographic entry is a refereed paper &
	N/A \\

\\
\noalign{\smallskip}
\hline
\end{tabular*}
\end{table*}

In the rest of this section we will discuss the approach we
followed in implementing the database structures 
allowing query and selection based on properties of bibliographies.
In section \ref{proprep} we describe the implementation used to associate 
properties and attributes to
entries in the database and the procedures maintaining relational
links among them.  In section \ref{propsoft} we describe the framework used to
automatically update and merge bibliographic data with information
submitted to the ADS.

\subsection {\label{proprep} Representation of Properties}

The creation and updating of properties in the ADS system 
is the result of merging entries provided by different data sources
and individuals at different times and in different formats.
The procedures used to maintain the property database are therefore
structured to be as general as possible (so that defining a new
property is a simple task) while still allowing as much customization
as necessary to deal with a variety of sources and formats.
The representation of properties allows the search engine to 
efficiently filter results based on whether a bibliographic entry
possesses a particular property.  It also allows fast access 
to the values associated to a particular bibliographic property, 
so that the search interface can quickly access the information
as required.

Instead of representing these properties as a single relational table
where each bibliographic entry is associated with the ordered set of 
property values, a different approach was chosen where each property 
is represented by a separate table.  
The following definition was adopted: 

``A bibliographic entry $b$ possesses property $p$ if the unique identifier
for $b$ appears in the property table associated with $p$, $T_p$.  
If $p$ is a property that can have
one or more values associated with it, the entry for $b$ in table $T_p$ will 
contain the $n$-tuple of such values next to it.''

As an example, a possible entry in table $T_{data}$
for a bibliographic entry which has a $data$ property associated to it
could be:

\begin{small}
\begin{verbatim}
1999A&A...341..121S
    http://cdsweb.u-strasbg.fr/htbin/myqcat3?
        J/A+A/341/121/
    http://adc.gsfc.nasa.gov/adc-cgi/cat.pl?
        /journal_tables/A+A/341/121/
\end{verbatim}
\end{small}

The first column contains the bibliographic identifier for the property,
while the second column contains the values of the $data$ property,
in this case a list of URLs of electronic data tables published in the
paper.  (Note that this record has been split on several lines for
editorial reasons.)

The file structure most amenable to representing these property tables
is again an inverted file, which allows fast binary searches on the 
bibcode identifiers.  As is the case for the inverted files used
to perform fielded searches on the contents of the bibliographic
entries in our database (see section \ref{indexing}), 
each property table is decomposed in two parts, an index file 
and a list file.  Since the records in the index file contain only
bibcodes, which have a fixed length, we can create a
binary index file where each record consist of one
bibcode identifier (which is the sort key in the file), a pointer 
into the list file, and the number of property values associated with 
the bibcode.
Entries in the list file are variable length, newline separated records,
each record corresponding to a property value.

In addition to the index and list files, 
a database-specific file is generated for each property
containing the list
of all bibcodes in that particular database which possess that
property.  
When the data structures used by the search engine are loaded into
random access memory, these lists of bibcodes are read and 
for each bibliographic entry a binary array containing the list 
of properties which it possesses is created.
By storing this information as part of the memory-resident data 
structures used by the search engine, selection and filtering of 
bibliographic entries based on their properties becomes a very efficient
operation.
The current implementation uses a 32-bit integer to represent the 
binary array of properties, where the $n$-th bit is set if and only
if the bibliographic entry possesses the $n$-th property.

\subsection {\label{propsoft} Implementation of the Property Database 
	Management Software}

\begin{figure}
\resizebox{\hsize}{!}{\includegraphics{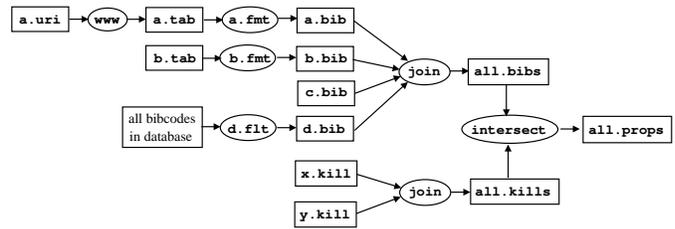}}
\caption[]{ Schema used for the creation of bibliographic
properties. In this abstract example, four different sources
contribute to the creation of bibliographic property files $a.bib,
b.bib, c.bib, d.bib.$ 
The input files used to generate the global list of properties may
consist of either static lists of bibcodes ($c.bib$),  tabular
data to be reprocessed to create properly formatted entries ($b.tab$),
lists of URLs containing information to be retrieved and  processed
($a.uri$), or ``filter'' functions acting on the global list of
bibliographic entries ($d.flt$). The system allows for the existence
of ``exception bibcodes,'' here represented as the contents of files
$x.kill$ and $y.kill$ that are removed from the global list of
bibcodes before the property inverted file $all.props$ is created. The
execution and updating of any of these files is controlled by a system
of makefiles that trigger updating only if necessary. }

\label{ADS_architectureF5}
\end{figure}

To provide the capability of merging properties and values generated
from separate sources and in different formats, 
we devised a framework consisting of a
hierarchical set of files and software utilities which 
are used to implement an efficient processing pipeline 
(see figure \ref{ADS_architectureF5}).  
The approach we follow may be regarded as being bottom-up, because
the property files are always created from smaller, independently
updated datasets.  Updating of such datasets is typically event-driven, 
as described below.

A top-level directory is created which
contains one subdirectory for each property in the database.  Each of
these subdirectories in turn contains files 
representing different datasets which need to be merged together.
The nature and content of such files is determined by their 
extension, according to the following conventions:

.tab: files containing identifiers and properties as provided by
	different data centers and users; these entries will need to
	be translated to the standard format used by scripts 
	managed by the ADS staff

.bib: files containing lists of tab-separated identifier and value pairs;
	these entries are suitable to be merged into a single property
	file used by the ADS search engine

.fmt: executable procedures which generate .bib files from their
	respective .tab files; these procedures contain
	format- and domain- specific knowledge about the source of the
	particular dataset and the mapping of entries from the .tab
	file into the .bib file

.uri: file containing the URLs of documents which should be downloaded
	from the network and merged to create a .tab file; these URLs
	may correspond to static or dynamic documents generated by
	other service providers listing the bibliographic properties
	available on their web site

.flt: executable procedures which generate .bib files by filtering
	the complete list of bibliographic identifiers according to
	some data-specific criteria; one example of such filter is the
	one which produces the list of all refereed bibcodes from the
	list of all bibcodes by checking the journal abbreviation

.kill: file containing the list of bibcodes which should $not$ be
	listed as possessing a particular property; these are
	typically used to implement ``exceptions to the rule,'' cases;
	for example, we use a kill file to remove bibcodes
	corresponding to editorial notices from the global list of
	papers appearing in a refereeed journal.

Data retrieval and formatting scripts designed after the GNU ``make''
utility limit the creation and processing of data to what is strictly
necessary.  
In particular, data sources that are specified as URLs are downloaded 
only if their timestamp is more recent than their local copy.
This obviously applies to network protocols that support the notion of
time-stamping, e.g. HTTP and FTP.
Similarly, scripts that are used to format input tables into lists of
bibcodes and relative URLs are only executed if the timestamp of the
relevant tables indicates that they have been modified more recently
than their corresponding target file.

\section {\label{mirrors} Database Mirroring}

All of the software development and data processing in the ADS has
been carried out over the last 6 years in a UNIX environment.
During the life of the project, the workgroup-class server used
to host the ADS services has been upgraded twice to meet the
increasing use of the system.  The original dual processor Sun 4/690
used at the inception of the project was replaced by a SparcServer
1000E with two 85MHz Supersparc CPU modules in 1995 and 
subsequently an Ultra Enterprise 450 with two 300MHz Ultrasparc CPUs
was purchased in 1997.  These two last machines are still currently
used to host the ADS article and abstract services, respectively.

Soon after after the inception of the article service in 1995 it
became clear that for most ADS users the limiting factor when
retrieving data from our computers was bandwidth rather than raw
processing power.
With the creation of the first mirror site hosted by the CDS in late 
1996, users in different parts of the world started being able 
to select the most convenient database server when using the ADS
services, making best use of bandwidth available to them.  
At the time of this writing, there are seven mirror sites located on
four different continents, and more institutions have already expressed
interest in hosting additional sites.
The administration of the increasing number of mirror sites requires a
scalable set of software tools which can be used by the ADS
staff to replicate and update the ADS services both in an
interactive and in an unsupervised fashion.

The cloning of our databases on remote sites has presented new
challenges to the ADS project, imposing
additional constraints on the organization and operation of our system.
In order to make it
possible to replicate a complex database system elsewhere, the
database management system and the underlying data sets have to be
independent of the local file structure, operating system, 
and hardware architecture.
Additionally, networked services which rely on
links with both internal and external web resources (possibly
available on different mirror sites) need to be capable
of deciding how the links should be created, giving users the
option to review and modify the system's linking strategy.  
Finally, a reliable
and efficient mechanism should be in place to allow unsupervised
database updates, especially for those applications involving the
publication of time-critical data.

In the next sections we describe the implementation of an efficient
model for the replication of our databases to the ADS mirror sites.
In section \ref{sysindep} we describe how system independence has been
achieved through the parameterization of site-specific variables and
the use of portable software tools.  In section \ref{siteindep} we
describe the approach we followed in abstracting the availability of
network resources through the implementation of user-selectable
preferences and the definition of site-specific default values.
In section \ref{mirrorsoft} we describe in more detail the paradigm
used to implement the synchronization of different parts of the ADS
databases.  We conclude with section \ref{mirrorenh} where we discuss
possible enhancements to the current design.

\subsection {\label{sysindep} System Independence}

The database management software and the search engine used by the
ADS bibliographic services have been written to be independent from
system-specific attributes to provide maximum flexibility in the 
choice of hardware and software in use on different mirror sites.
We are currently supporting the following hardware architectures:
Sparc/Solaris, Alpha/Tru64 (formerly Digital Unix), IBM RS6000/AIX,
and x86/Linux.  Given the current trends in hardware and operating
systems, we expect to standardize to GNU/Linux systems in the future.

Hardware independence was made possible by writing portable
software that can be either compiled under a standard compiler and
environment framework (e.g. the GNU programming tools, \cite{GNU})
or interpreted by a standard language (e.g. PERL version 5,
\cite{PERL}).  Under this scheme,
the software used by the ADS mirrors is first compiled from a common
source tree for the different hardware platforms
on the main ADS server, and then the
appropriate binary distributions are mirrored to the remote sites.

One aspect of our databases which is affected by the specific
server hardware is the use of binary data in the list files, since
binary integer representations depend on the native byte ordering
supported by the hardware.
With the
introduction of a mirror site running Digital UNIX in the summer of
1999, we were faced with having to decide whether it was better to
start maintaining two versions of the binary data files used
in our indices or if the two integer implementations should be handled
in software.  While we have chosen to perform the integer conversion
in software for the time being given the adequate speed of the
hardware in use, we may revisit the issue if the number of mirror
sites with different byte ordering increases with time.

Operating System independence is achieved by using a standard
set of public domain tools abiding to well-defined POSIX standards
(\cite{POSIX}).  Any additional enhancements to the
standard software tools provided by the local operating system
is achieved by cloning more advanced
software utilities (e.g. the GNU shell-utils package) and using them
as necessary.
Specific operating system settings which control kernel parameters 
are modified when appropriate to increase system performance and/or
compatibility among different operating systems 
(e.g. the parameters controlling access to the system's shared memory).
This is usually an operation that needs to be done only once
when a new mirror site is configured.

File-system independence is made possible by organizing the data
files for a specific database under a single directory tree, and
creating configuration files with parameters pointing to the location
of these top-level directories.  Similarly, host name
independence is achieved by storing the host names of ADS servers in
a set of configuration files.

\subsection {\label{siteindep} Site Independence}

While the creation of the ADS mirror sites makes it virtually 
impossible for users to notice any difference when accessing the
bibliographic databases on different
sites, the network topology of a mirror site and its connectivity
with the rest of the Internet play an important role in the way
external resources are linked to and from the ADS services.  With the
proliferation of mirror sites for several networked services in
the field of astronomy and electronic publishing, the capability to
create hyperlinks to resources external to the ADS based on the
individual user's network connectivity has become an important issue.

\begin{figure}
\resizebox{\hsize}{!}{\includegraphics{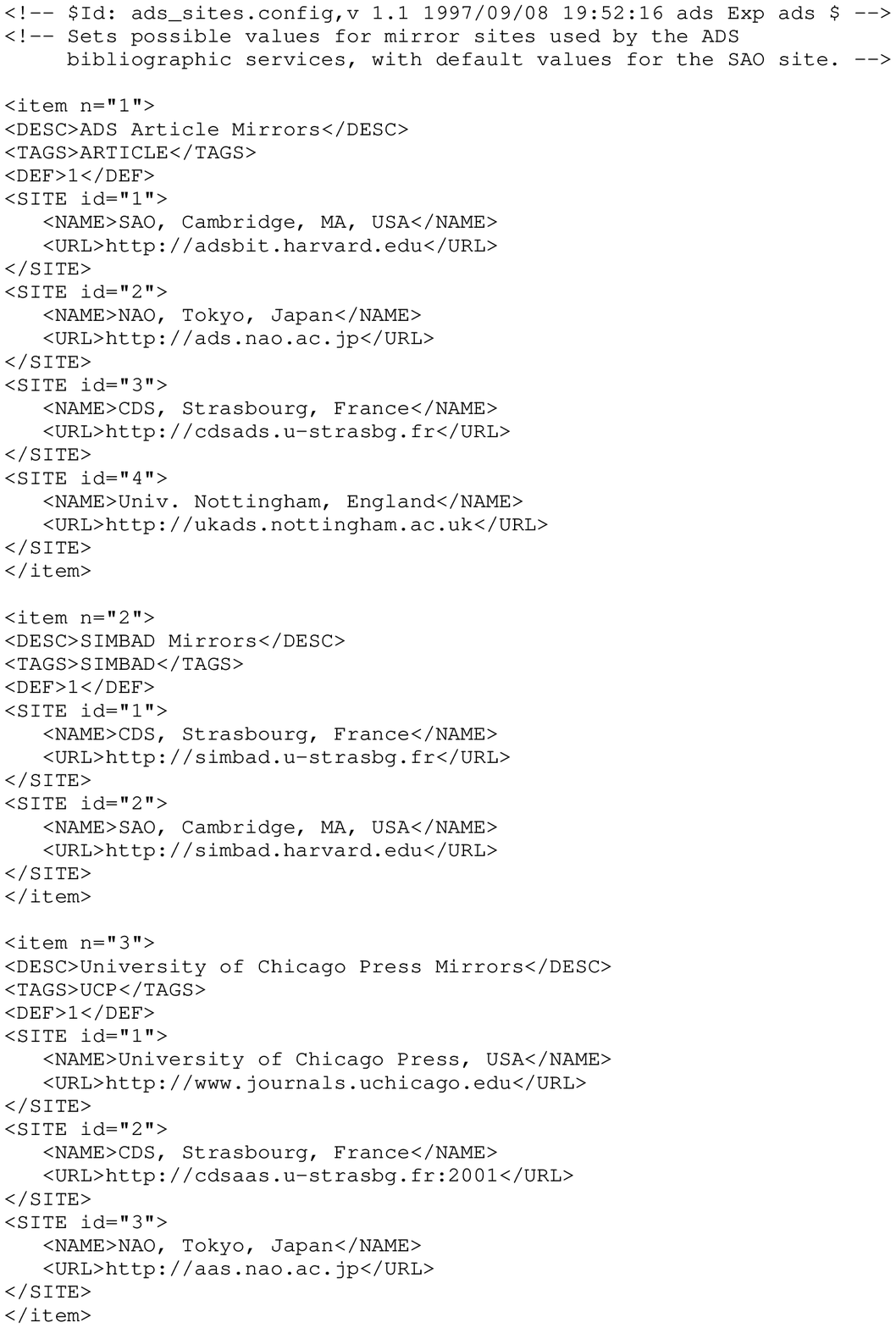}}
\caption[]{ The configuration file used to define variables and
related sites for resources available from multiple network
locations. It should be noted that this approach can be used for
parameterizing and generalizing URL resolution even in those cases
where the resource is available from a single location. }

\label{ADS_architectureF6}
\end{figure}

The strategy used to generate links to networked services external to
the ADS which are available on more than one site follows a
two-tiered approach.  First, a ``default'' mirror can be specified in a
configuration file by the ADS administrator (see figure  \ref{ADS_architectureF6}).
The configuration file defines a set of parameters used to compose
URLs for different classes of resources, lists all the
possible values that these parameters may assume, and then defines a
default value for each parameter.  Since these configuration files are
site-specific, the appropriate defaults can be chosen for each of the
ADS mirror sites depending on their location.
ADS users are then allowed to override these defaults by
using the ``Preference Settings'' system (SEARCH) to
select any of the resources listed under a category as their default
one.  Their selection is stored in a site-specific user preference
database which uses an HTTP cookie as an ID correlating users with
their preferences (SEARCH).

In order to create links to external resources which are a function of
a user's preferences, we store the parametrized version of their URLs in
the property databases.  The search engine expands the
parameter when the resource is requested by a user
according to the user's preferences.  
For instance, the parametrized URL for the
electronic paper associated with the bibliographic entry 
{\tt 1997ApJ...486...42G} can be expressed as 
{\tt \$UCP\$/cgi-bin/resolve?1997ApJ...486...42G}.  Assuming the user has
selected the first entry as the default server for this resource, the
search engine will expand the URL to the expression:
\begin{verbatim}
http://www.journals.uchicago.edu/cgi-bin/resolve?
    1997ApJ...486...42G
\end{verbatim}
This effectively allows us to implement simple name resolution for a
variety of resources that we link to.
While more sophisticated ways to create dynamic links have been
proposed and are being used by other institutions 
(\cite{VDS99,FERNIQUE98}), there is currently no
reliable way to automatically choose the ``best'' mirror site for a
particular user, since this depends on the connectivity between the
user and the external resource rather than the connectivity
between the the ADS mirror site and the resource.
By saving these settings in a user preference
database indexed on the user HTTP cookie ID (SEARCH), 
users only need to define their preferences once and our interface 
will retrieve and use the appropriate settings as necessary.

\subsection {\label{mirrorsoft} Mirroring software}

\begin{figure}
\resizebox{\hsize}{!}{\includegraphics{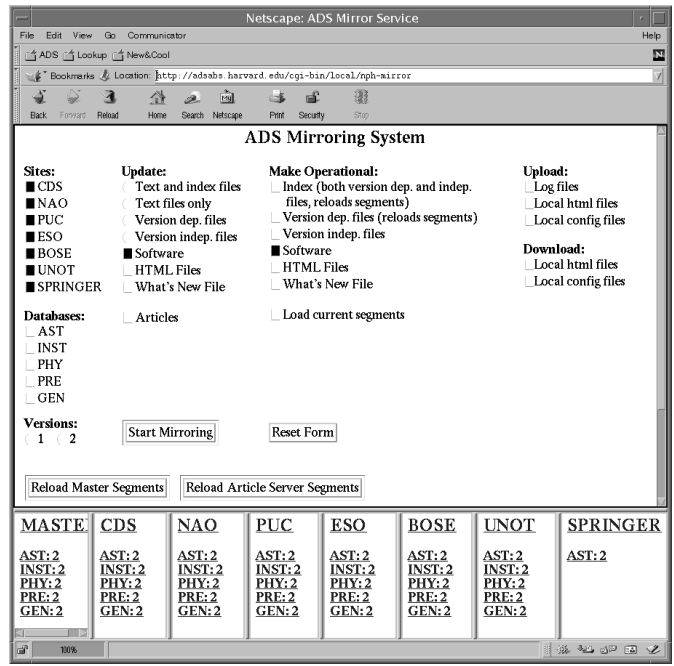}}
\caption[]{ The WWW interface used by the ADS administrators to update
different components of the databases on the different mirror
sites. The small windows at the bottom of the screen display, for each
of the ADS databases, the version number currently operational at each
mirror site. }

\label{ADS_architectureF7}
\end{figure}

The software used to perform the actual mirroring of the databases
consists of a main program running on the ADS master site initiating
the mirroring procedure, and a number of scripts, run on the mirror
sites, which perform the transfer of files and software necessary to
update the database.  
The paradigm we adopted in creating the tools used to maintain the
mirror sites in sync is based on a ``push'' approach: updates are
always started on the ADS main site.  
This allows mirroring to be easily controlled by the
ADS administrator and enables us to implement event-triggered updating of
the databases.
The main mirroring program, which can be run either from
the command line or through the Common Gateway Interface (CGI), 
is a script that initiates a remote shell session on the remote sites
to be updated, sets up the environment by  evaluating the mirror sites' and
master site's configuration files, and then runs scripts on the remote
sites that synchronize the local datasets with the ADS main site.
An example of the menu-driven CGI interface and a mirroring
session are shown in figure \ref{ADS_architectureF7}.

The updating procedures are specialized scripts which check and
update different parts of the database and database management
software (including the procedures themselves).  
For each component of the database that needs to be updated,
synchronization takes place in two steps, namely the remote updating
of files which have changed to a staging directory, and the action of
making these new files operational.  This separation of mirroring
procedures has allowed us to enforce the proper checks on integrity
and consistency of a data set before it is made operational.

The actual comparison and data transfer for each of the files to be
updated is done by using a public domain implementation
of the rsync algorithm (\cite{TRIDGELL99}).
The advantages of using rsync to update data files rather than
using more traditional data replication packages are summarized below.

1) Incremental updates: rsync updates individual files by scanning
  their contents, computing and comparing checksums on blocks of data
  within them, and copying across the network only those blocks that
  differ.   Since during our updates only a small part
  of the data files actually changes,
  this has proven to be a great advantage.  Recent implementations of the 
  rsync algorithms also allow partial transfer of files,
  which we found useful when transferring the large index files 
  used by the search engine.  In case the network connection is lost
  or times out while a large file is transferred, the partial file
  is kept on the receiving side so that transfer of additional 
  chunks of that file can continue where it left off on the next
  invocation of rsync.

2) Data integrity: rsync provides several options that can be used to
  decide whether a file needs updating without having to compare its
  contents byte by byte.  The default behavior is to initiate a block
  by block comparison only if there is a difference in the basic file 
  attributes (time stamp and file size).  The program however can be
  forced to perform a file integrity check by also requesting a match
  on the 128-bit MD4 checksum for the files.

3) Data compression: rsync supports internal compression of the data
  stream sent between the master and mirror hosts by using the zlib
  library (\cite{ZLIB}). 

4) Encryption and authentication: rsync can be used in conjunction with the 
  Secure Shell package (\cite{SSH}) to enforce authentication
  between rsync client and server host and to transfer the data in an
  encrypted way for added security.
  Unfortunately, since all of the ADS mirror sites are outside of the
  U.S., transfer of encrypted data could not be performed at
  this time due to restrictions and regulations on
  the use of encryption technology.

5) Access control: the use of rsync allows
  the remote mirror sites to retrieve data from the master ADS site
  using the so-called anonymous rsync protocol.  This allows the
  master site to exercise significant control over which hosts are
  allowed to access the rsync server, what datasets can be 
  mirrored, and does not require remote shell access to the main ADS
  site, which has always been the source of great security problems.

During a typical weekly update of the ADS astronomy database, as many
as 1\% of the text files may be added or updated, while the index files
are completely recreated.  By checking the attributes of the
individual files and transferring only the ones for which either 
timestamp or size has changed, the actual data which gets transferred
when updating the collection of text files
is of the order of 1.7\% of the total file size (12MB vs. 700MB).
By using the incremental update features of rsync when mirroring a new
set of index files, the total amount of data being transferred is
of the order of 38\% (250MB vs. 650MB).

\subsection {\label{mirrorenh} Planned Enhancements}

While the adoption of the rsync protocol has made it possible to
dramatically decrease the time required to update a remote database,
there are several areas where additional improvements could be made to
the current scheme in an effort to reduce the amount of redundant
processing and network transfers on the main ADS server.  Some of the
planned improvements are discussed below.

Given the CPU-intensive activity of computing lists of file signatures
and checksums for files selected as potential targets for a transfer,
the rsync server running on the main ADS site is often under a heavy
load when the weekly updates of our bibliographic databases are
simultaneously mirrored to the remote sites.  
Under the current implementation of the rsync server software, each
request from a mirror site is handled by a separate process which
creates the list of files and directories being checked.
Therefore, the load on the server increases linearly with the
number of remote hosts being updated, although much of the processing
requested by the separate rsync connections is in common and takes
place at the same time.
By adding an option to cache the data signatures generated by the
rsync server and exchanged with each client, most of the processing
involved could be avoided.  This option, first suggested by the author
of the rsync package (\cite{RCACHE}) but never implemented, would
significantly benefit busy sites such as the ADS main host.
A similar approach has been used by \cite{DEMPSEY99} to implement an
experimental replication mechanism based on rsync.  We hope that a
stable and general approach to this caching issue can be adopted soon
and are collaborating with the maintainers of the package on its
development. 

\begin{figure}
\resizebox{\hsize}{!}{\includegraphics{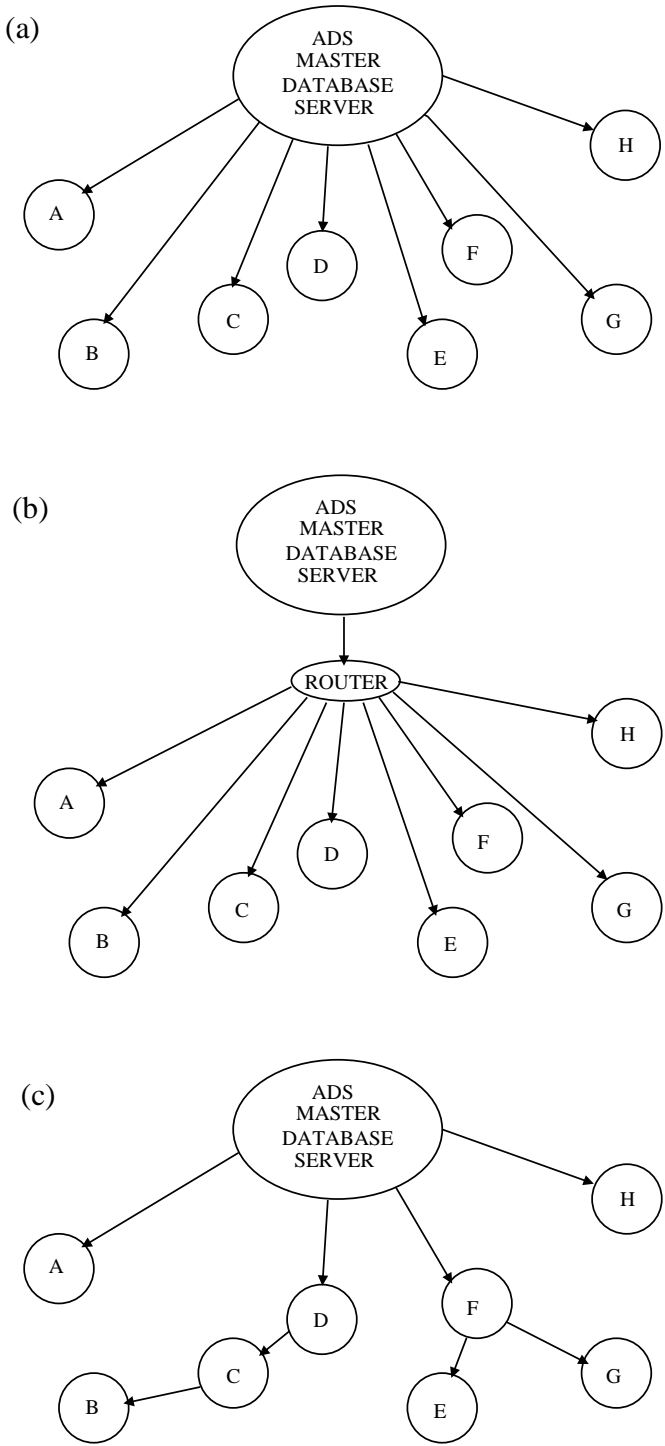}}
\caption[]{Schematic representation of network mirroring models used 
to replicate a central database available on the ADS master database 
server to a number of mirror sites (A-H).  Traditional model (a): data 
is transferred using parallel, independent data pipes between the master 
and the mirror sites. Multicasting model (b): one single stream of data 
is transferred  from the master site to a central router and then 
multiplexed to the mirror sites using multicast technologies. Cascading 
model (c): a hierarchy of mirror nodes is defined based on the relative 
network connectivity; each node updates the local copy of its 
databases and then proceeds to mirror them to its subordinate nodes. }

\label{ADS_architectureF8}
\end{figure}

A second improvement that would significantly reduce the bandwidth
currently used during remote updating of the ADS mirror sites is the
implementation of a multicasting or cascading mirroring model (see
figure \ref{ADS_architectureF8}).  Internet multicasting is still a
technology under development (\cite{MILLER98}) and efficient
implementations require special software support at the IP (Internet
Protocol) level, over which we have no control.  
The cascading model can instead by implemented at the application
level using current software tools.  Under this model, the
administrator of the main server to be cloned defines a tree in which
the nodes represent the mirror sites, with the root of the tree being
the main site.  Data mirroring is then implemented by having each node
in the tree ``push'' data to its subordinate nodes.
This approach trades off the simplicity of simultaneous updating 
for all mirror sites from a central host
in favor of a sequence of cascading updates, which is a sensible
solution once the number of mirror sites becomes large.
We are currently experimenting with this model on a prototype system and
plan to make the design operational by the end of 1999 if the design
proves to be advantageous.

\section {\label{discussion} Future Developments}

By all accounts, the ADS project has been very successful in providing
bibliographical services to the astronomer and research librarian.
Much of the system's strength has been its role as part 
of a network of services designed to provide advanced search and retrieval
capabilities to the scientific community at large.
Given the rapid changes in the field of electronic publishing,
resource linking, digital library research,
it is of great importance for our project to adapt its operations
to this ever-changing environment and its underlying technologies.

In this last section we analyze some of the promises and challenges
that we expect to face over the next several years and we discuss
how they may affect the evolution of our system.
In section \ref{newdata} we describe the new datasets that are
becoming available to our project and the changes necessary for their
integration in the existing system architecture.
Section \ref{newtech} describes the effect of expected 
technological changes on the operations of the ADS.
Finally, section \ref{newserv} discussed how increased collaboration
and inter-operability among data providers can lead to the creation of
a more integrated environment making better use of 
information discovery and electronic publishing technologies.

\subsection {\label{newdata} New Data}

From the user prospective, one of the most significant changes will be 
the completion of our full-text coverage and abstracting for the 
scholarly astronomical literature.  
Over the next year we expect to complete the digitization of all
astronomical journals back to volume 1 (DATA).
The availability of such a large body of scanned publications
allows us to pursue some important goals through the use 
of Optical Character Recognition (OCR) technology: the creation of
full-text documents and the extraction of abstract and citation 
information from them.

The full text of an article produced by OCR programs can be used
by the indexing and search engine to provide better retrieval
capabilities.  However, the current indexing model has been
developed to work well with a homogeneous set of bibliographic
data with little variation in document length and content model;
extending the scope of our databases to include the full-text of
articles may therefore require a new approach to the entire 
architecture behind the indexing and search engines.
Furthermore, since the output generated by OCR packages is known to
contain incorrectly recognized characters and words, new strategies
may be required to manage this level of uncertainty during indexing
and searching.

The extraction and OCRing of important document fragments such as
abstracts and references is currently an ongoing process which holds
great promise (DATA).
Essentially, the combination of pattern recognition and OCR 
techniques allows us to identify areas in a scanned document
corresponding to the abstract or reference section of a paper.
The text extracted from an abstract section is then reformatted
and inserted into the bibliographic record for that paper.
Periodic analysis of the text index has been 
necessary to identify and correct misinterpreted characters
and words produced by the OCR software.
The increased amount of human checks on our data set as a quality
assurance measure has been the price to pay for integrating these
additional abstracts in our bibliographic records.

Text extracted from a reference section is analyzed by 
programs making use of natural language processing techniques 
to identify the individual works cited in the article and
add them to our citation database.
The challenge we are facing in this case is creating a robust system
capable of correctly parsing and matching the cited reference
strings with bibliographic records in our database
(\cite{1999adass...8..291A}), with the additional complication that the
input text may contain characters incorrectly recognized by the
OCR software.

\subsection {\label{newtech} New Technologies}

The latest developments in Electronic Data Interchange and User 
Interfaces advocate the adoption of a model of data
representation where there is clear separation between content,
metadata, and style.
The widespread endorsement of XML and related proposals such as
the XLink language, the Extensible Style Language (XSL), and 
the Document Object Model (DOM), seems to indicate that we will 
see pervasive use of XML across platforms and implementations.
While this raises hopes that data exchange among different
astronomical data centers and institutions can be streamlined, it is not clear 
at this point that a unique framework describing all resources
in astronomy can be defined, nor that such a system is necessary
at this point.
However, the adoption of XML as the ``lingua franca'' for data 
interchange can help remove the initial obstacles preventing more
widespread creation of peer-to-peer connections between information
providers and can help speed up the creation 
of ``federated'' services (\cite{AML}).

In this context, we hope to leverage the wide deployment of XML-based
applications to generalize and extend the services currently offered 
to our collaborators and users.
This involves modifying the implemented APIs (SEARCH) to
allow output of structured XML documents containing both metadata and 
bibliographic data. 
We have already started adopting this paradigm while implementing
new and experimental services which require the exchange of data and
metadata structures between client and server, 
such as the ADS reference resolver
(\cite{1999adass...8..291A}).

Another issue related to data interchange which is currently 
receiving much attention is the definition of persistent identifiers
for bibliographic resources available on the Internet.  This issue
is a particular instance of a more general problem, which is 
the need to define common naming schemes for digital objects 
and distributed locator
services allowing their resolution.
For a number of years this has been recognized as one of the 
most important infrastructure components necessary for the 
large-scale development of digital library systems (\cite{DLWR}).
Today most publishers are providing
location services which are based on the traditional paradigm of
identifying a published work by journal, volume and page.
It is becoming increasingly clear that a more general mechanism
will have to be adopted in the future since this model does not
extend well into the digital era.
For instance, a publication may be available only in electronic
form (as is already the case for some ``e-journals'' such as
EPJdirect and ZPhys-e from Springer-Verlag).
or may correspond to a multimedia object rather than a traditional
text document;  in these cases, the concept of pagination loses
its meaning. 
The Document Object Identifier (DOI, \cite{DOI}), which has been
proposed by an international consortium of publishers, holds the 
promise of becoming the universal identifier suitable for 
naming digital objects.  Unfortunately, the required registration
procedures and management of DOI space and limited support for its
location services seem to have discouraged
its widespread adoption so far (\cite{DAVIDSON98}).

The ADS has already extended the use 
of the bibcode identifier in different ways to account for the 
existence of electronic-only publications (DATA), but it is 
becoming increasingly more difficult to map new document
identifiers into a model that was designed to describe printed
material only.
It is likely that over the next few years our project will need
to adopt new notations for identifying bibliographic records, 
while still maintaining backward compatibility with the existing
bibcodes for printed work.
In this sense, it is likely that ADS will be able to help the
astronomical community in the transition from print-based to
electronic publishing by providing resolving services for astronomical
bibliographies and related resources.

\subsection {\label{newserv} New Services}

The adoption of common technologies and protocols by data providers has
helped create a low-level of inter-operability among different
data services (in the sense that users can simply browse across different
web sites by following links between them).
However, with the exponential increase of documents and
services available on the web, the problem of providing an 
integrated tool for locating information of interest to a 
researcher has remained unsolved.  While well-organized 
repositories and archives with good search interfaces 
exist for a variety of data sets, a scientist who needs to
consult several such archives is left with having to
individually query each one separately and then organize
the results collected from each one of them.
It is fortunate that the creation of the ADS and its ongoing
collaboration with other data providers has reduced (if not completely
eliminated) this problem for astronomers, but this is not the case for
scientists in other disciplines or for those researches whose work
spans across the conventional boundaries of scientific research fields.

The problem of providing a unified search 
mechanism across datasets is being tackled both within the
individual disciplines (\cite{1999adass...8..221H,FERNIQUE98,AML}) 
and at the architectural level (\cite{SCHATZ97}).
A proposed solution to this problem is the creation of federated services
composed by ``clustering'' the combined assets and search capabilities
of several independent data centers.  
A common set of metadata elements 
describing the local search domain and interface can be used to
translate generic queries into site-specific ones, and then
merge and present the results to the users.
While this type of approach is known to work within well-restricted
research domains, the broader problem of querying databases belonging
to different research fields is far more complex and requires
the creation of systems capable of implementing semantic
inter-operability (\cite{SCHATZ97,DLWR}).
While the ADS has been offering direct access to its search engine
since 1996 (SEARCH), 
in order for the ADS to become part of such a federated system, we
will need to provide an increased level of abstraction and access to
the capabilities of our search interfaces.  
Additionally, the emerging standards for site- and 
database-specific resource descriptions will require the creation and
maintenance of a body of metadata defining both the extent of our
databases and the supported query interfaces.
\cite{HANISCH99} has recently proposed the creation of such a
distributed system for Astronomy and the Space Sciences.

Another important aspect of services increasing inter-operability
between data providers is cross-linking of online resources.
While most publishers of scientific journals have been able to 
create electronic versions of their journals relatively quickly
soon after the explosion in popularity of the web, only a few of them
have taken advantage of the new capabilities that the technology
has to offer, namely the possibility to create hyperlinks between
online documents and related resources.
In this respect, electronic publishing in astronomy was ahead of 
its times with the publication by the University of Chicago Press
in late 1996 of the electronic version of the Astrophysical Journal
which contained hyperlinks from the reference section of
articles to bibliographic records in the ADS.
The early implementation of this feature became possible thanks
to the close collaboration between the publisher, the ADS staff,
and the visionary leadership provided by the American Astronomical
Society (AAS).
Similarly, editors and publishers have now made it their policy to
submit electronic versions of data tables appearing in astronomical
papers to the CDS and Astronomical Data Center (ADC) archives,
allowing ADS to easily maintain links to these datasets in its
bibliographic records.  This practice was estabilished back in 1990
with an agreement between the CDS and the editors of the journal
Astronomy \& Astrophysics. 

While reference and object linking has today become more commonplace
(\cite{HITCHCOCK98}), there are a number of unresolved problems that
limit its usefulness.
The issue of linking a reference to an instance of the document it
refers to can be viewed as a two step process (\cite{CAPLAN99}):
(1) resolution of a reference string into a document identifier; 
and (2) resolution of the document identifier into one or more
URLs.
In the current use of the ADS reference resolver, 
(\cite{1999adass...8..291A}) step (1) is 
accomplished by the publisher during the last stages of the electronic
publication process, and links are created only if a reference string
is found to correspond to a valid bibcode in ADS (``static linking'').
The step of document resolution (2) is another example of the problem
of object resolution mentioned in section \ref{newtech}.  In this case,
a bibcode needs to be mapped into the ``best'' URL corresponding to
it, and is typically implemented as a site-specific resolution
activity, so that for example, the CDS mirror of the University of
Chicago journals will link to the CDS mirror of the ADS bibliographic
services.

While this model has worked well for many astronomical journals,
it has some shortcomings.
First of all, the computation of static links at publication time
does not allow for the possibility that one of the works cited in
the reference section may become available at a later date (e.g.
if the coverage of the literature has been extended or if a more
accurate resolution of the reference is later implemented).
From a theoretical point of view, a better approach to the problem
would be the use of ``dynamic linking,'' in which links are created
when the document is downloaded (\cite{VDS99}).
It is likely that most publishers will move towards a mixed model 
in which on-line documents are periodically reprocessed for the
purpose of updating links between them and external resources
that may have become available, or to provide options for
forward-looking citation queries into bibliographical databases.

As far as the issue of bibcode resolution, it is clear that a better
approach to having site-specific settings would be to allow real-time
resolution of bibcode identifiers based on the preference of the
individual users and the current availability of relevant resources.
The approach we follow when resolving links to external resources
(SEARCH) does account for user preferences, but does not take into
account real-time availability of the possible instances of the
resource. 
This is in contrast with the approach followed by \cite{FERNIQUE98}, where
the opposite is true.
It is clear that in order to create a reliable system for resolving
astronomical resources, and integration of both approaches is
necessary, so that a global user profile can be used to specify
preferences while a global resource database can be used to specify
the availability and location of these resources on the network.
The implementation of such a system is greatly complicated by the
increasingly complex organization of networks, with firewalls
and proxy servers acting as intermediary agents in the activity of
resource resolution.  
Hopefully these issues will be solved over the next few years by the
adoption of standard practices and software tools.

\section {\label{conclusions} Conclusions}

The design and implementation of the ADS bibliographic services has
been driven by the desire to provide flexible search capabilities to
the astronomical community. 
The original decision to create our own suite of software tools for
indexing and searching the databases has proven to be an important one
as it has given us the freedom to continuously enhance and tailor the
software to our users' needs.
With freedom, however, also came the responsibility of maintaining a
complex system which has now been ported to a variety of hardware and
software platforms.  Fortunately,
the adoption of standard programming languages and coding techniques
has greatly facilitated the task.

Over the years, the ADS has evolved from being a user-oriented
system to becoming an open service for the discovery and
retrieval of bibliographic data, allowing integration of our
capabilities in the operation of other information providers.
At the same time, our system was expanded from being simply a
searchable archive of bibliographic references to being a
service offering relational links among records within our system and 
to resources available elsewhere.
In this respect, the design of a hierarchical framework for the
management of bibliographic resources has provided the required
level of flexibility and extensibility.
With the recent proliferation of mirror sites for popular resources in
astronomy, we have adopted a simple yet powerful mechanism for the
resolution of links to resources available at multiple
locations, adding user customization to the
resolution process. 

With the completion of full-text coverage of the astronomical literature
over the next few years, the ADS will be able to significantly increase
the holdings of its citation database and provide full-text search and
retrieval capabilities.
With the adoption of new technologies and standards in electronic
data interchange, the ADS is likely continue to play an important role in the
integration of network services in astronomy.

\acknowledgements

The usefulness of a bibliographic service is only as good as
the quality and quantity of the data it contains.  The ADS project
has been lucky in benefitting from the skills and dedication
of several people who have significantly contributed to the 
creation and management of the underlying datasets.
In particular, we would like to acknowledge the work of 
Elizabeth Bohlen, Donna Thompson, Markus Demleitner,
and Joyce Watson.

Funding for this project has been provided by NASA under grant NCC5-189.

\end{document}